\documentclass[fleqn,10pt]{wlscirep}
\usepackage[utf8]{inputenc}
\usepackage[T1]{fontenc}

\usepackage[percent]{overpic} 

\title{Mid-career pitfall of consecutive success in science}

\author[1,+*]{Noriyuki Higashide}
\author[1,+]{Takahiro Miura}
\author[1]{Yuta Tomokiyo}
\author[1]{Kimitaka Asatani}
\author[1]{Ichiro Sakata}

\affil[1]{Department of Technology Management for Innovation, Graduate School of Engineering, The University of Tokyo, Tokyo 113-8656, Japan}
\affil[+]{These authors contributed equally to this work}
\affil[*]{n.higashide@ipr-ctr.t.u-tokyo.ac.jp}

\begin{abstract}
The creativity of scientists often manifests as localized hot streaks of significant success. Understanding the underlying mechanisms of these influential phases can enhance the effectiveness of support systems and funding allocation, fostering groundbreaking discoveries worthy of accolades. Historically, analyses have suggested that hot streaks occur randomly over time. However, our research, through meticulous examination, reveals that these phases are not flatly distributed but are more frequent at the early and late stages of scientists’ careers. Notably, both early and late hot streaks are marked by dense tie collaborations, with the former typically involving close partnerships with particular authors and the latter being characterized by involvement in large-scale projects compared with single-top or ordinary papers. This pattern indicates that mid-career researchers lack both intimate relations and resources to keep big projects, leading to ``mid-career pitfall'' of consecutive success. This insight holds profound implications for the development of policies and initiatives aimed at bolstering innovative research and discovery.
\end{abstract}

\begin{document}

\flushbottom
\maketitle
%
%
\thispagestyle{empty}



\section*{Introduction}

Over the past half-century, the role of scientists has expanded beyond simply writing quality papers, making it more difficult to focus solely on academic interests due to the growing need for broader engagement in academia.
University faculty members are increasingly busy with committee meetings for selecting other faculty members and Ph.D. candidates, grant reports and presentations, teaching, meetings with lab team members, trips to discuss with external collaborators, and attending a kick-off event for a new research consortium\cite{lashuel2020busy}. 
This diversification of a scientist's skill set dilutes the resources available for accelerating sciences\cite{chu2021slowed}, even emerging the possibility that the labor advantage in scientific fields could be becoming a bottleneck in driving productivity\cite{zhang2022labor}.
Given the escalating challenges in achieving remarkable results with limited time and resources, it is important to understand the mechanisms that enable scientists to be actively engaged in their work from a more systematic and ecosystem-based perspective. 
This understanding is crucial for research support, including those involved in research management initiatives, funding, and science and technology policy, as it poses a significant challenge that needs to be clearly addressed.

The phenomenon known as hot streaks\cite{liu2018hot,liu2021understanding} is one of the distinguishing markers of a researcher's engagement.
Einstein's ``miracle year,'' where he made four revolutionary discoveries, are typical example of a hot streak. 
New ideas often do not arise in isolation but are part of a chain where one discovery sparks another. 
Previous research shows that scientists' top three papers are typically produced in a consecutive sequence\cite{liu2018hot}, so scientific success somewhat manifests with a curious continuity.
Triggers for consecutive success may be more valuable than triggers for a singular success because single success may merely represent the fortunate extraction of a good idea from a universal distribution of ideas\cite{sinatora2016quantify}, whereas the hot streak can be seen as a `bonus time' that elevates the impact of every output over a certain period\cite{liu2018hot}.
Thus, uncovering the universal patterns of hot streaks can help efficiently boost researchers' intellectual creativity, potentially revolutionizing how we stimulate and sustain scientific innovation.

In investigating the mechanisms that could trigger hot streaks, a type of career success, the career ages should be considered\cite{jones2014age}.
Hot streaks are supposed to occur with a certain probability throughout any career ages\cite{liu2018hot}, but if we assume this legitimacy, consistently publishing consecutive high-impact works during the mid-career stage seems quite challenging due to the large number of involved projects, heavy workloads, funding issues, and other factors that demand attention.
These challenges are explained in Nature Career's special podcast series \textit{``Muddle of the Middle''}\cite{muddle2022}. 
In their mid-career, having recently earned their Ph.D. and within the first few years of obtaining tenure, face a unique set of challenges compared to their younger counterparts who often enjoy more funding and support, or even harder than senior scientists who have settled assets and experienced skills to manage various activities. 
These mid-career researchers are required to juggle an array of responsibilities, including managing their first teams with limited budgets, preparing new lectures, and consistently publishing papers to secure their next positions. 
Consequently, mid-career researchers may find it difficult to allocate enough time for both exploration and exploitation which often marks the onsets of hot streaks\cite{liu2021understanding}.

It's valuable to examine how hot streaks, \textit{i.e.} consecutive successes, of scientists differ by career stages.
Scientists' academic career successes are related to their collaboration patterns such as team formations and co-authoring\cite{petersen2015quantifying, zeng2019increasing, wu2019large, li2019early, ma2020mentorship, zeng2021fresh, yang2022gender, li2022untangling, liu2022team, zeng2022impactful, lin2023remote}.
Hot streaks are more likely to be produced by larger teams\cite{liu2021understanding}, but team composition approaches vary among career stages. 
Early-career researchers might be likely to focus on building fresh connections while working closely with their supervisors\cite{li2019early, ma2020mentorship, zeng2021fresh}. 
In contrast, senior researchers might leverage their established networks with previous collaborators and effectively manage super-ties, ensuring long-term collaboration\cite{petersen2015quantifying, zeng2022impactful}. 
What kind of collaboration produces consecutive success?

In this study, we investigated when top papers are produced in a researcher's career and what factors are behind their creation depending on career stages, using a large-scale literature database. 
Contrary to previous studies, our results revealed a U-shaped pattern, indicating that hot streaks are more common in the early and later stages of a career.
For the early and later stages, we clarified the differences in supporting factors from the perspectives of team collaboration.

\section*{Consecutive successes concentrate on a career's early and late stages}

\begin{figure}[bth]
    \centering
    \includegraphics[width=\linewidth]{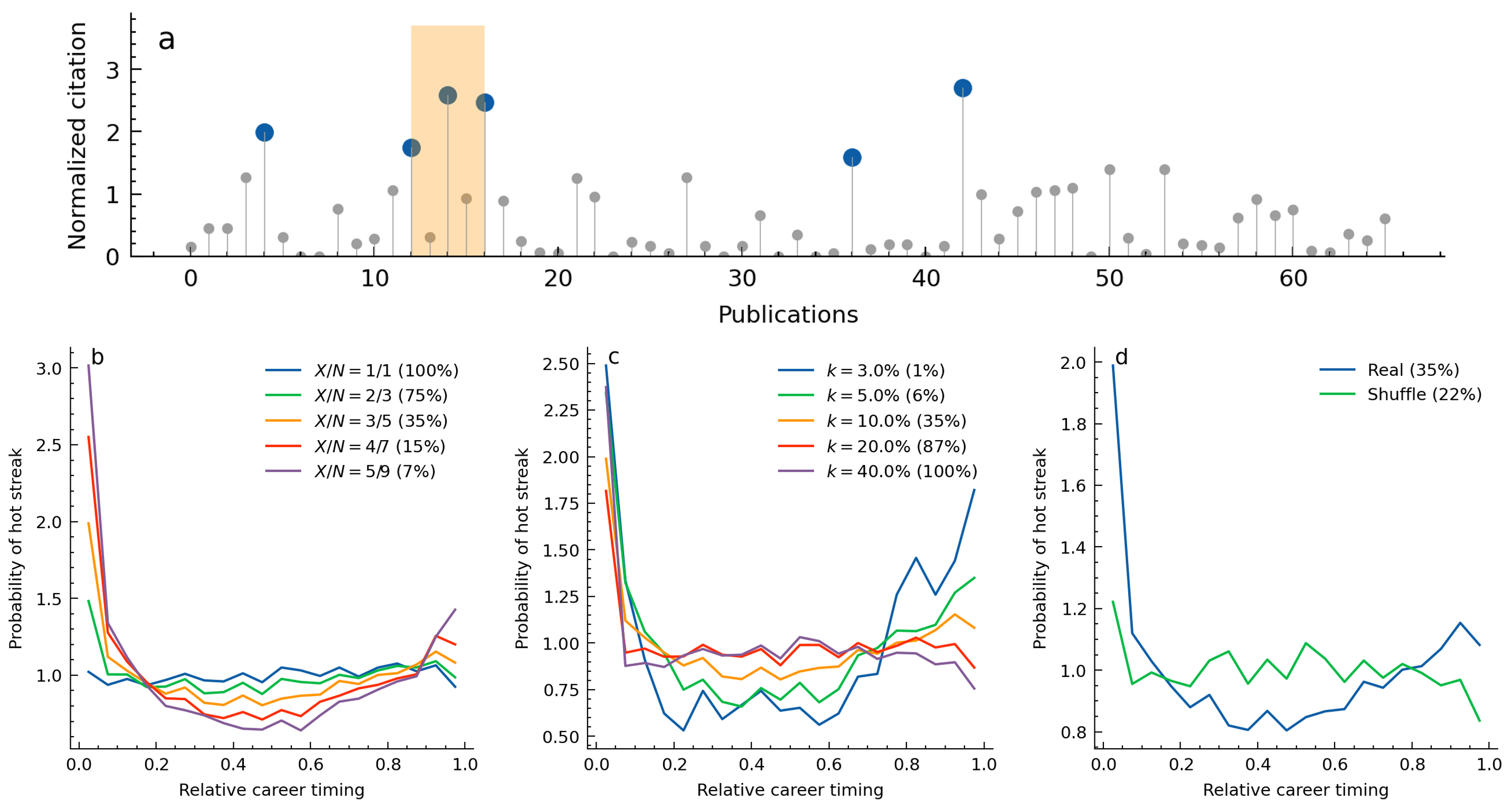}
    \caption{(a) Definition of consecutive success. Top $k$\% impactful papers are shown as blue circles and they appear $X$ times within $N$ publications with parameters $X=3, N=5, k=10\%$. We use the number of publications in sequence as the length of the researchers’ career instead of years. (b-d) Probability distribution when consecutive success occurs in the scientists' careers. The x-axis, relative timing, represents normalized career length. The numbers in parentheses represent the percentages of researchers, out of 100,000, who have experienced at least one hot streak under that parameter. As consecutive success becomes stronger, the number of researchers experiencing consecutive success decreases. (b) The distributions of consecutive success with different length-related parameters $X$ and $N$ ($k=10\%$). A single success ($X/N=1/1$) occurs with a constant probability but long continuous successes ($X/N=5/9$) occur more frequently in the early and late throughout the career. (c) The distributions of consecutive success for different $k$ ($X/N=3/5$). Big success occurs more in the early and late stages. (d) compares the distribution between the raw and shuffled career sequences with parameters $X/N=3/5$ and $k=10\%$.}
    \label{fig:u-shape}
\end{figure}

A researcher's career history of publication impact is used to identify consecutive success. The career sequence dataset was crafted from over 100,000 scientists with careers spanning more than 20 years and 30 publications (Supplementary information \ref{sec_sup:data}). 
We defined a consecutive success as a duration when scientists’ top-$k$\% most impactful papers are concentrated $X$ times within $N$ publications (see \hyperref[sec:methods]{Methods} for mathematical descriptions). 
In Figure \ref{fig:u-shape}a, for example, the top $10\%$ high-impact works indicated by blue circles, are concentrated with $X=3$ out of $N=5$ works in the period from 12 to 16 marked in orange which represents a consecutive success period.
If these periods are detected consecutively within overlapping windows, these are consolidated and considered as a single consecutive success with a longer length. Various parameters can be set to detect consecutive successes. A larger $N$ indicates longer consecutive successes, while a smaller $N$ means shorter ones. 
Instances detected with parameters like $X/N=5/9$ are regarded as strong hot streaks, and $X/N=1/1$ captures the weakest form of ones, which is a single success.
The occurence timing of consecutive success is the starting position of consecutive $N$ papars. The timing is normalized for different career lengths among researchers. For example, the relative timing of the $i$-th paper in a career is calculated as $i/(N_{T}-N)$, where $N_{T}$ is the total number of published papers.

The previous study captured consecutive success by using the hot-streak model\cite{liu2018hot}. 
The hot-streak model assumes a temporal rise in a researcher's potential and thus is applied for the moving average of career sequence data with the window size of $\Delta N = max(5,0.1N_T)$. The size is ceiled by 10\% of a career length to smooth out influence by random factors.
However, we find that the window size could determine the duration of hot streaks (Supplementary information \ref{subsec_sup:hs-model-limits}). The average duration of 3.7 years derived by the hot-streak model may be an artifact based on single successes stretched by the window size, suggesting it may not accurately capture the true sequence of consecutive successes in a career.
Contrarily, our approach more directly characterizes consecutive successes by examining raw career sequence data with fewer artificial assumptions.

Detected consecutive successes with several parameters show intriguing characteristics. Firstly, we found that these are less likely to occur in the mid-stage of a career but are more likely to appear in the early and late stages (Fig. \ref{fig:u-shape}b-d). 
Especially, single success are spread flat in the career, but consecutive successes concentrate on career beginnings and ends (Fig. \ref{fig:u-shape}b). 
Further, moderate successes, $k=50\%$ for example, happen randomly, but intensive successes like $k=5\%$ are more likely to happen in the early and later phases (Fig. \ref{fig:u-shape}c). 
Shuffling the order of career sequence data eliminates the U-shaped distribution, which means that the pattern of early and late career successes is not a coincidental phenomenon (Fig. \ref{fig:u-shape}d). 
These findings indicate that consecutive successes do not occur at a constant probability throughout a career. 
Instead, the likelihood of their occurrence is concentrated on the early and late stages, especially for successes of greater length and intensity. 

Looking at the starting year distribution of our consecutive success extraction, the observed years of consecutive success tend to increase as careers progress into their later stages. However, compared to instances of single hits, the mid-career periods, from 7 years after the career begins to 5-10 years before the career ends, show lower likelihoods of achieving consecutive hits (Supplementary information \ref{subsec_sup:probability_year}). 
This illustrates the dynamics whereby true consecutive successes manifest initially during the Ph.D. and early postdoctoral phases, subsequently subside during the mid-career period, and become less likely until the final stages of one's career.
These U-shaped patterns demonstrate robustness over time (Supplementary information \ref{subsec_sup:probability_time}) and, even when considering variations in career dynamics of success across different fields, mostly hold true (Supplementary information \ref{subsec_sup:probability_field}).
Our identification method shows U-shaped distributions over the career but the hot-streak model demonstrates that consecutive successes occur with consistent probability throughout careers\cite{liu2018hot}. 
We estimated that this discrepancy mainly comes from preprocessing for career sequence data and false positive detection of the previous method (Supplementary information \ref{sec_sup:liu_model}).

\section*{Patterns behind consecutive successes}

\begin{figure}[h!tb]
    \centering
    \includegraphics[width=\linewidth]{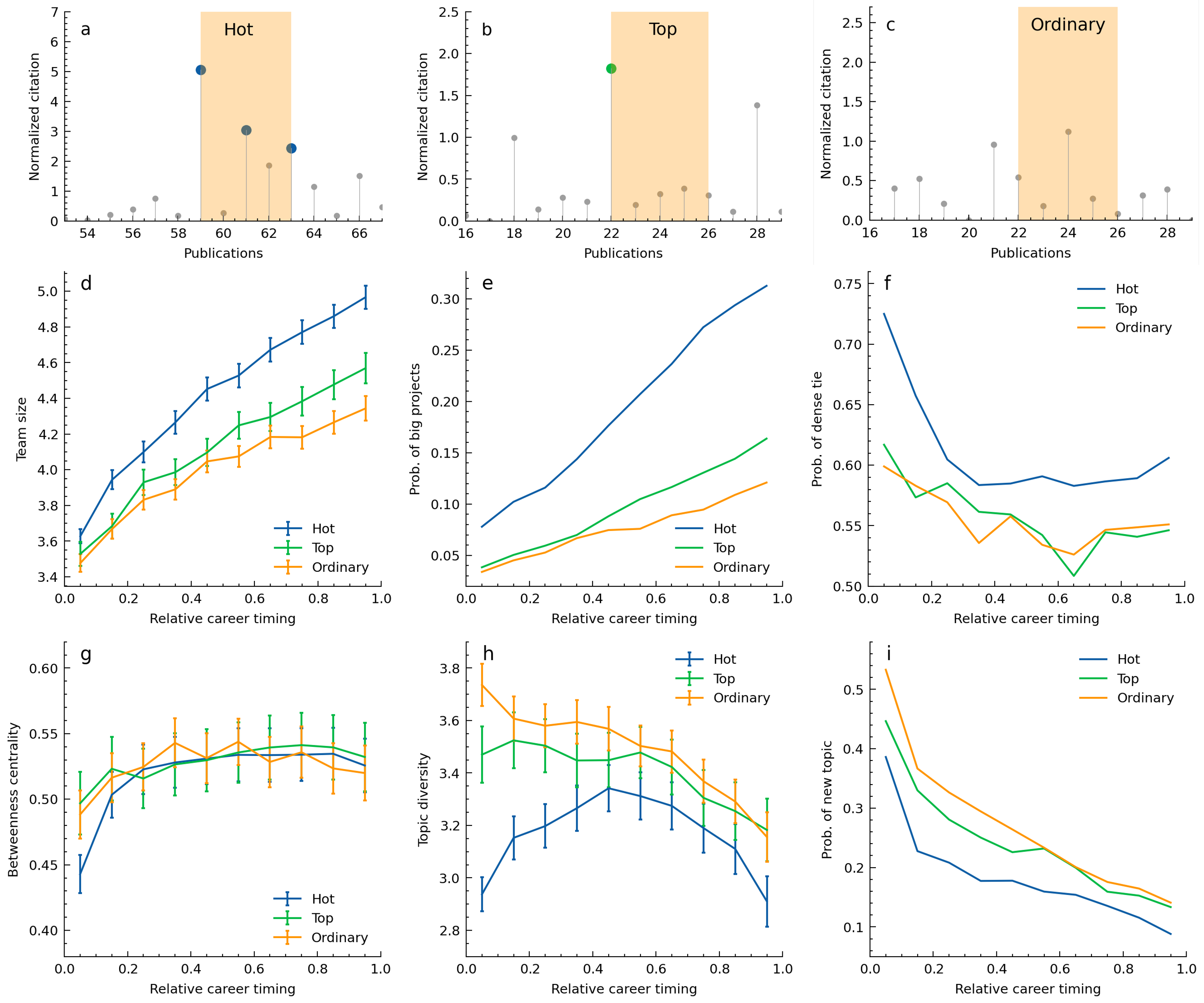}
    \caption{Characteristics of consecutive success. (a) A typical example of a `Hot' career sequence. The period filled in orange is detected as a hot streak of five consecutive papers. Blue dots represent top 10\% papers, appearing three times within the five papers. (b) An example of `Top' sequence, represents a period of single success, not consecutive. A green dot indicates a top 10\% paper, appearing once among five papers. (c) `Ordinary' sequence example as comparisons, five consecutive non-top 10\% papers, representing a period without significant hits in careers. (d-i) Comparison of the 3 sequences across 6 metrics. The relative career timing is divided into ten bins, and metrics are calculated for sequences with the same start timing in each bin. This allows observation of how metrics vary in early, mid, and late career stages. (d) Team size, focused on author lists of fewer than 10 people (Supplementary information \ref{sec_sup:team}). (e) Proportion of teams with 10+ members. (f) Proportion of sequences with dense ties, appearing 3 or more times in the co-author list of the 5 consecutive papers. (g) Betweenness centrality of the focal author in the co-authorship network of the 5 papers. (h) The number of topics tackled in the 5 papers, ranging from 1 to 5. (i) Proportion of new topics not previously tackled. In (d) and (f-i), calculations are performed on sequences where the team size for all 5 papers is less than 10. The bars in (d,g,h) represent the 95\% confidence interval that includes the estimated population mean, assuming a normal distribution. See \hyperref[sec:methods]{Methods} for details of definitions and calculations of each metric.}
    \label{fig:characterization}
\end{figure}


To understand factors related to the occurrence of consecutive successes, we characterize the career-stage dependency of indicators related to team activities.
Science and collaborations are closely intertwined, with factors such as team size\cite{wu2019large}, the strength of connections between authors\cite{petersen2015quantifying, li2022untangling}, and the topics\cite{zeng2019increasing, zeng2022impactful} particularly influencing scientific outcomes.
Based on these indicators, our findings revealed unique patterns associated with consecutive successes.

We defined three types of career sequences: `Hot', `Top', and `Ordinary' (Fig. \ref{fig:characterization}a-c). `Hot' represents consecutive success with parameters of $X=3$, $N=5$, $k=10\%$, that is five consecutive papers by the same author including at least three highly-cited ones. `Top' denotes single success, including five consecutive works starting with a top 10\% paper but fewer than three high-impact papers. `Ordinary' works comprise five non-top 10\% papers. Our analysis covered a sufficient number of each across all career stages: 47,356 `Hot', 25,227 `Top', and 39,278 `Ordinary' sequences (see \hyperref[sec:methods]{Methods}). This comparison helps us to uncover several patterns of `Hot' career sequences.

1. \textbf{Large teams}: We found that larger teams are more common in later career stages and for `Hot' sequences, suggesting that consecutive successes are more likely to arise from larger teams than `Top', and `Ordinary' (Fig. \ref{fig:characterization}d), which is consistent with 
the previous study\cite{liu2021understanding}. 
This trend is measured in teams of nine or fewer members.

2. \textbf{Big projects}: We defined teams of 10 or more as 'big projects' and calculated the probability of such projects. If at least one paper in a consecutive period is a big project, the sequence is counted as a big project.
The later career stages have a higher probability of big projects, with `Hot' sequences notably more prevalent than the other two types (Fig. \ref{fig:characterization}e), indicating that big projects are likely to be in a series of high-impact works of an individual's career.
The probability of `Hot' sequences having papers with more than 100 authors is about ten times higher than the other two types, no matter the career stage (\ref{fig_sup:teamsize_prob}).

3. \textbf{Dense ties}: To understand collaboration patterns in depth, we analyzed co-authorship networks of five consecutive papers (\hyperref[sec:methods]{Methods}). 
Observing the presence of dense ties, defined as co-authoring three or more of the five papers with the same authors, we found that `Hot' sequences consistently show a higher probability of dense connections throughout a career (Fig. \ref{fig:characterization}f). 
We also hypothesized that authors with consecutive successes might have higher betweenness centrality in co-authorship networks, acting as hubs connecting different author groups. 
However, no significant overall difference is observed, suggesting that the network structure of collaborations doesn't vary significantly (Fig. \ref{fig:characterization}g). 
A higher collaboration rate with specific authors could contribute to experiencing successful periods.

4. \textbf{Topic focus}: The diversity of research topics in five consecutive papers is investigated (\hyperref[sec:methods]{Methods}). 
`Hot' sequences generally exhibited less topic diversity, indicating a more focused approach and the pattern shows an inverse U-shape with a decrease in early stages, an increase in mid-career, and a decrease in later stages (Fig. \ref{fig:characterization}h). 
The stronger focus in early and late stages might relate to the distribution of consecutive successes being more common in these phases. 

5. \textbf{New topics}: We also examined the probability of new topics tackled during the sequences, which had not been addressed in the author's previous career (\hyperref[sec:methods]{Methods}). 
In `Hot' sequences, throughout a career, the likelihood of venturing into new topics is lower than in the other two types (Fig. \ref{fig:characterization}i).
This indicates that revisiting previously tackled research topics is related to consecutive success, providing hints for topic selection strategy.

Overall, consecutive successes tend to occur in larger teams, with frequent collaboration with the same authors, focusing on fewer topics, and these topics are often previously addressed. These patterns suggest particular strategies and environments facilitating successive high-impact scientific works at different career stages.

\section*{Early consecutive success from dense ties, later from large teams}

What explains the U-shape distribution of consecutive successes along careers? In the early stages, we see small teams, dense ties, and a focus on research topics. These suggest that early-career researchers experience hot streaks with working intensively with a few co-authors, such as mentors and their group members. As careers progress to the middle stage, the number of collaborators increases, the probability of working with specific individuals decreases, and topic diversity grows. This trend might reflect the expanding collaborator network over time. In later stages, team size grows further and researchers tend to focus on specific research topics. The established reputation and authority of senior scientists might enable them to steer large teams toward focused topics.

\begin{figure}[h!tb]
    \centering
    \includegraphics[width=\linewidth]{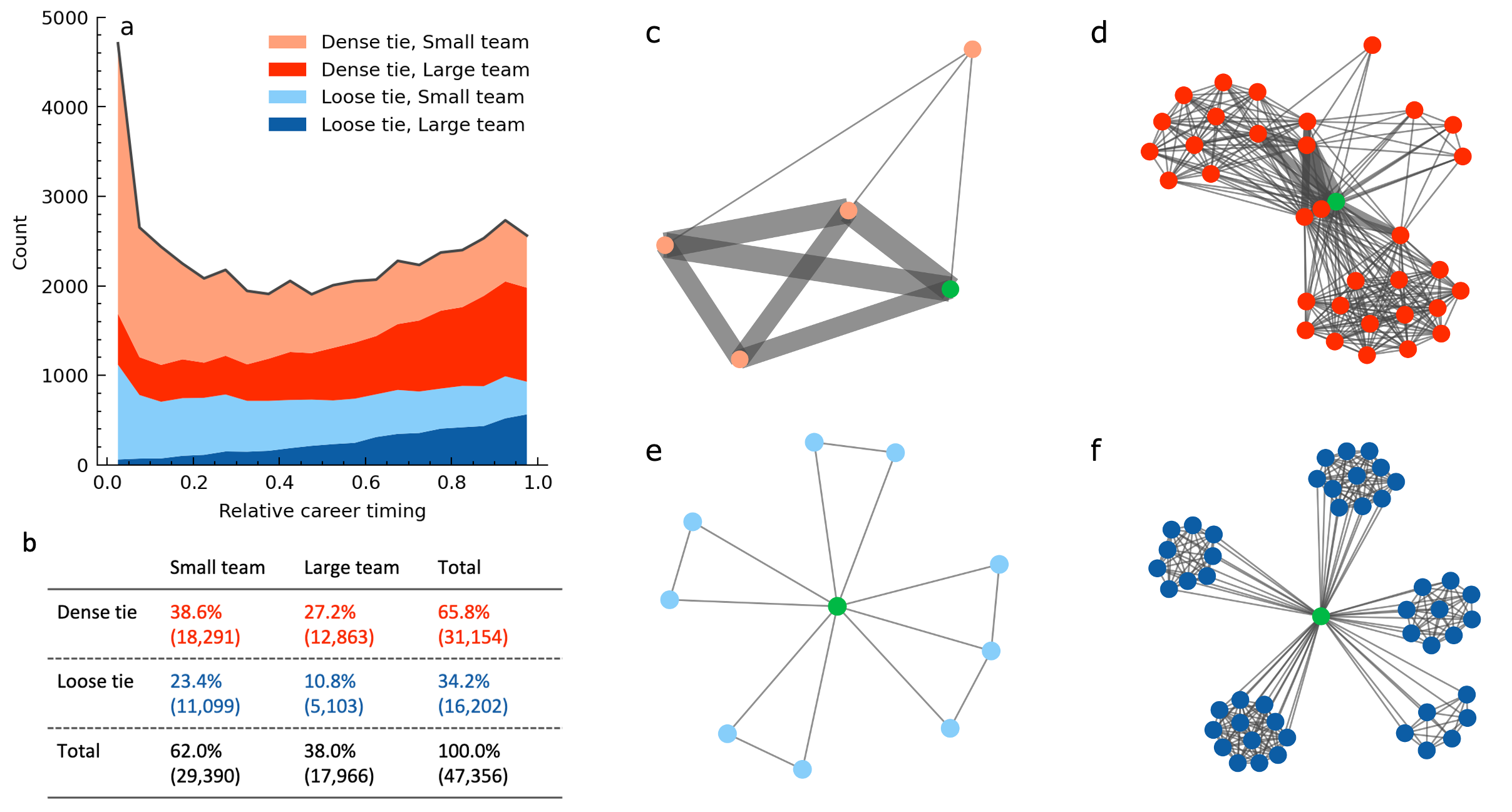}
    \caption{Four types of consecutive successes. (a) The histogram when consecutive successes occur with parameters $X/N=3/5$ and $k=10\%$. The four types of them are displayed in a stacked manner, each represented by a different color. The top two types, shown in pink and red, represent consecutive successes with dense ties, more common in early and late stages. The bottom two, in light blue and blue, depict loose ties occurring consistently across the career. (b) The four types' proportions. 65\% of them originated from dense ties and 40\% originated from large teams. (c-f) Typical co-authorship networks for each type. Green nodes represent the focal author, other nodes are co-authors, and edges indicate the number of collaborations in five consecutive papers, ranging from one to five. (c) Dense ties in small teams are shown as thick edges between a few nodes. (d) Thick edges with large teams composed of many co-authors. (e) The focal author is connected with fewer authors and thinner edges representing loose ties and small teams. (f) No thick edges with many co-authors show collaboration in loose ties with large teams.}
    \label{fig:explanation}
\end{figure}

Given the distributions of consecutive successes involving dense ties and big projects (Fig. \ref{fig:characterization}e, f), we find that these two factors contribute to the U-shape distribution shown in Figure \ref{fig:u-shape}b. 
Consecutive successes within loose ties, which do not involve as frequent collaboration as dense ties, occur with almost constant probability throughout a career, while those involving dense ties contribute to the spike at the early stages and the slight increase in the later stages (Fig. \ref{fig:explanation}a). 
As careers progress into the later stages, the proportion of consecutive success originating in small teams decreases, while those arising from larger teams increase. Hereinafter, small teams mean less than 10 members and large teams mean ten or more (\textit{i.e.} big projects).
These indicate that both early and late career successes are dominated by dense ties, while later stages see stronger dominance by larger teams.
Thus, by considering two factors, the strength of connections and team size, we can categorize types of consecutive successes into four distinct groups.
Notably, about 40\% of consecutive successes originate from teams of more than ten members (Fig. \ref{fig:explanation}b). 
Each type is intuitively understood by visualizing their co-authorship networks (Fig. \ref{fig:explanation}c-f).


Dense ties and large teams can explain why mid-career scientists less frequently experience consecutive successes. During mid-career, dense connections are fewer than in the early stages, and the team size is not sufficiently large. This situation could make consecutive successes more challenging.

\begin{figure}[h!tb]
    \centering
    \includegraphics[width=\linewidth]{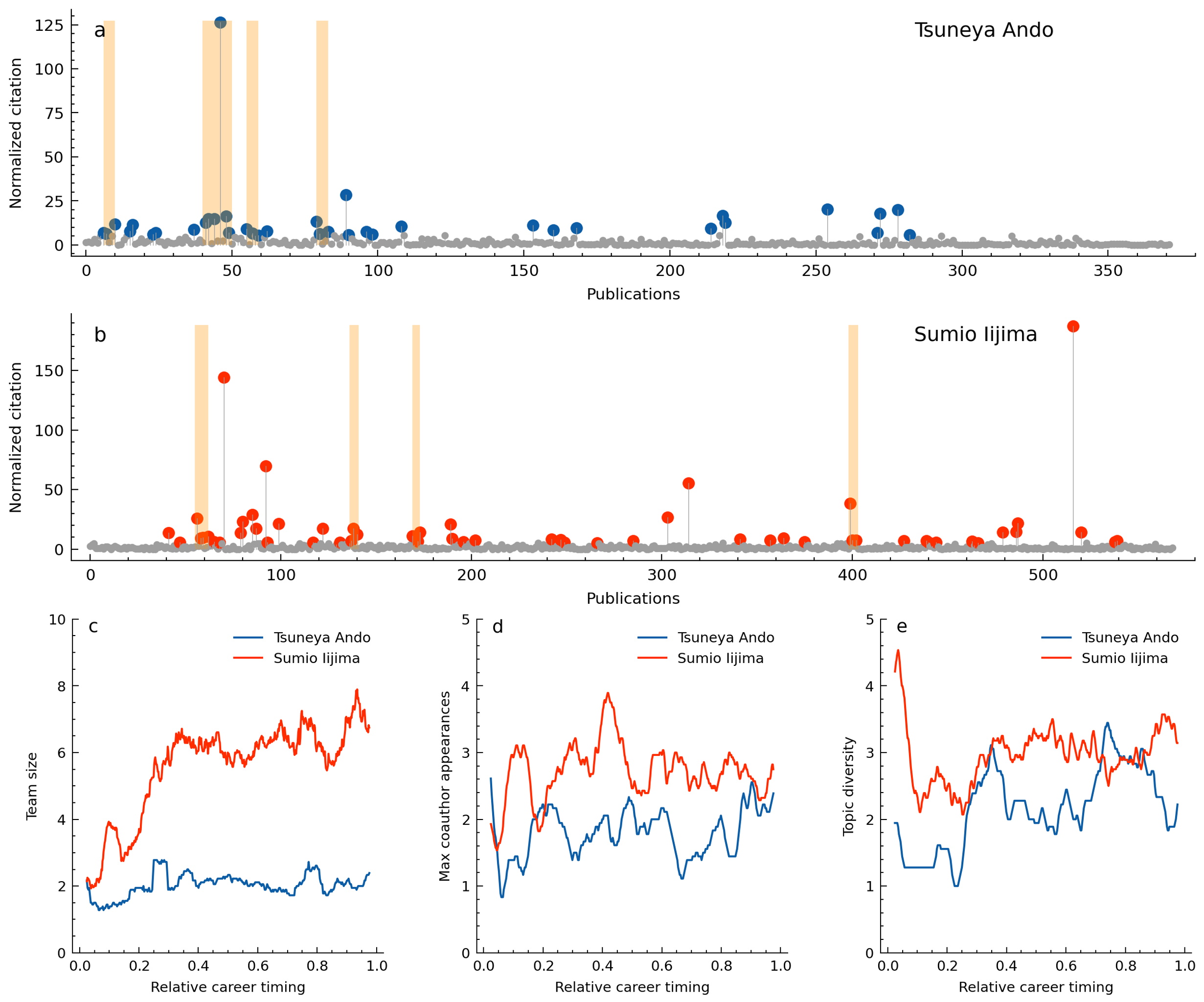}
    \caption{Two scientists' typical career dynamics. (a,b) Career sequences. The papers with top-10\% impact are colored and the orange areas indicate consecutive success identified by our method. (c) Relation between papers' publication timing in normalized career length and their team size for two scientists. (d) Frequency of co-authors in five consecutive papers; higher values mean repeated collaborations with the same authors, reflecting dense ties. (e) Topic diversity dependency on career stages, measured by the variety of topics in five consecutive papers. For clarity, (c-e) display moving averages with a window size of 5\% career length.}
    \label{fig:hs_example}
\end{figure}

As examples, we consider two physicists in condensed matter physics. The field has received significant contributions from Japanese scientists, including some renowned researchers. Tsuneya Ando made significant theoretical contributions, predicting the quantum Hall effect in semiconductors. Sumio Iijima is a great experimental researcher who discovered carbon nanotubes. Both have produced over 350 papers throughout their careers, showing their high productivity.
Ando produced high-impact works early in his career, going through four consecutive successes (Fig. \ref{fig:hs_example}a), with his most significant work included in his second one. On the other hand, Iijima had three consecutive successes early in his career and then another later on (Fig. \ref{fig:hs_example}b). Tracing their work trajectories revealed interesting differences. Upon normalizing the length of their careers and comparing team size, collaboration density, and topic diversity, it was found that Iijima worked with larger teams later in his career (Fig. \ref{fig:hs_example}c). The maximum co-author frequency and topic diversity across five papers tend to be higher for Iijima throughout his career, but the difference is relatively small (Fig. \ref{fig:hs_example}d,e). Indeed, in the later stages of his career, Iijima produced his most significant paper in graphene, outperforming his early discoveries, with over 10 co-authors. Such a pattern of consecutive success with larger team sizes in the later career aligns with the overall trend observed among scientists (Fig. \ref{fig:characterization}d).


\section*{Discussions}

In this study, we hypothesized that the mechanisms underlying consecutive success vary depending on career stages and thus focused on analysis from the perspective of collaboration patterns. 
The hot-streak model\cite{liu2018hot} may not always capture continuous success because, in the process of applying moving averages, there's a possibility that single successes may be mistakenly identified as consecutive successes (Supplementary information \ref{subsec_sup:hs-model-limits}). Extracting consecutive successes by more directly observing sequences of highly-cited works, we found that a series of successes is more likely to happen in the early and later stages of a career, with a dip during mid-career. 
Early-career hot streaks are more likely among researchers with dense ties to specific colleagues, whereas late-career hot streaks are often supported by big projects on familiar research topics. 

The uncovered empirical regularities highlight the importance of considering a scientist successes at each career stages.
Early in the career, dense connections are present, but as the career progresses and the academic network expands forward, larger team sizes emerge, showing a complementary relationship. The ``mid-career pitfall'' may be attributed to the loss of these two factors.

The mid-career, spanning from the 10th to 20th year, is challenging. Increased independence and responsibility, departures from initial support structures, and the expansion of administrative tasks requiring multitasking make it difficult for researchers to focus on specific topics and allocate resources. 
Researchers not only move between organizations seeking positions\cite{vaccario2021reproducing} but also belong to more research communities, frequently switching topics\cite{zeng2019increasing}.
These career dynamics increase the cost of adapting to new environments. 
In addition to these factors, a decline in the number of publications in mid-career\cite{sabharwal2013comparing}, looks influence mid-career dip of consecutive success. 
The discontinuous success in mid-career may be due to limited time, human resources, and the difficulty of returning to familiar settings. Yet, in the later stages, forming large teams can contribute to a revival of hot streaks. 
Taken together, this suggests that if a stable research foundation is established earlier in the mid-career, it would lead to greater achievements which could further accelerate scientific progress.

The reality of mid-career researchers' support is also reflected in national research support policies, such as the Japan Society for the Promotion of Science, which divides funding opportunities into categories open to all and those exclusively for young researchers\cite{jsps2024}. 
Consequently, mid-career researchers are forced to compete with more experienced researchers for funding, potentially influencing the occurrence of hot streaks. 
Single successes can occur randomly at any career stage, suggesting that such kind of success can be achieved even without funding, but sustaining successes and maximizing output on specific topics might require new funding strategies targeted at the overlooked mid-career researchers.

The prevalence of hot streaks in the early stages of a career suggests a potential survivorship bias within the scientific community. 
Our study, focusing on researchers with at least 30 publications over a 20-year period, indicates that those who peak in the first five years of their career are likely to sustain their research career with ease. 
However, this approach may overlook hidden talents not represented in our data. 
The emergence of hot streaks among researchers with dense ties, particularly in the early stages, implies that survival as a scientist may depend not only on individual potential but also on the chance of forming a hot streak through good mentorship initially. 
Prior research on the positive effect of connecting with top researchers early in one's career \cite{li2019early} suggests that this not only impacts the magnitude of later success but also the likelihood of sustaining a research career.

The divergence of success distribution due to exploration, followed by a re-concentration towards the end of a career, is highly suggestive. 
Researchers in their later career stages shine by forming large teams of more than ten people and refocusing on familiar topics. 
This is particularly evident in experimental fields that can expand the scale of experiments with significant budgets, such as Health Science and Manufacturing Engineering, where many hot streaks are experienced in the last 10\% of a career (Fig. \ref{fig_sup:ushape_field}). 
In contrast, fields emphasizing theory, like Algebraic Geometry and Network Science, experience fewer late hot streaks. 
This disparity reflects the relationship between funding strategy and fields where costs and outcomes are more or less directly proportional. In experimental fields, forming large teams can lead to hot streaks at any age.

Our study has limitations common to this style of analysis.
Firstly, the results may be biased due to the dataset used. Since Scopus is limited to reputable journals designated by Elsevier, the observed consecutive successes are confined to works of a certain level of renown. 
By restricting data to the years 1970 to 2012, the dataset may not fully capture the beginning or end of the careers of individuals who started their careers before 1970 or whose careers continued beyond 2012 (Supplementary information \ref{subsec_sup:limitation_data}).
Secondly, while this study focuses on common trends in how collaboration patterns change during hot streaks across careers, there also is the diversity across disciplines (Fig. \ref{fig_sup:ushape_field}).
The U-shape pattern does not appear universally across all fields, indicating that a more detailed analysis, delving into the content of the disciplines, is necessary in the future to understand what determines the occurrence of hot streaks. The patterns discovered in this study specifically target only academic careers and may not directly apply to a wider range of creative fields, such as artists, film directors,  musicians, or ballet dancers\cite{liu2021understanding, oliveira2023hot, herrera2023quantifying}.
Lastly, the findings of this study represent correlations rather than causations. 
While it is true that dense ties in the early stages and larger teams in the later stages are commonly observed during current hot streaks, real experiments are required to determine whether interventions can lead to the occurrence of hot streaks.

\section*{Methods}
\phantomsection
\label{sec:methods}
\textbf{Detection of consecutive successes.}
Given an author \(a\), the set of papers published by this author is denoted as \(P_a = \{p_1, p_2, ..., p_{N_T}\}\), where $N_T$ is the total number of published papers.
The 10-year normalized citation count for a paper \(p\) is represented by \(C_{10}^p\).
The subset of papers \(P_a^{\text{k}} \subset P_a\) includes papers whose \(C_{10}^p\) are in the top \(k\%\) of all \(C_{10}^p\) in \(P_a\).
For a specified window \(N\), a time interval from \(i\) to \(i+N\) is considered as a consequtive success if:

\begin{equation}
\left|\{p \in P_a^{\text{k}}\  |\  p \in p_i, ..., p_{i+N} \}\right| \geq X \label{eq:our}
\end{equation}

Here, \(|\cdot|\) denotes the cardinality of a set. Eq. (\ref{eq:our}) indicates that if the number of papers in \(P_a^{\text{k}}\) published within the window from \(i\) to \(i+N\) is more than \(X\), the interval to be considered consequtive success.
This method detects overlapping periods in a career sequence, and we have incorporated a process to merge such instances. As a result, the length of consecutive success is usually five papers, but it can be longer. The number of people experiencing hot streaks and the frequency of these occurrences vary depending on the definition (Fig. \ref{fig:u-shape}b-c). 
Compared with previous hot-streak model\cite{liu2018hot,liu2021understanding}, which capture the potential change of reseacher's output, this definition is more data-centric approach focusing on the actual output of researchers.

Some authors experience more than one hot streak during their careers. Under the conditions of $X = 3, N = 5, k = 10\%$, this proportion is 13\%. We also found that authors who produce more papers over their careers tend to experience hot streaks more frequently, and those who have hot streaks generally have a higher overall career impact. When handling career sequence data, we use the number of publications instead of years to normalize career patterns that vary across fields due to different publication frequencies and numbers\cite{sinatora2016quantify}. 

\textbf{Three types of sequences.} To compare patterns of consecutive success across different career stages, we defined `Hot', `Top', and `Ordinary' sequences (Fig. \ref{fig:characterization}a-c). To determine these, first, we identified 47,356 hot streaks from 35,495 authors with the parameters $X = 3$, $N = 5$, and $k = 10\%$. Some hot streaks were longer than five papers due to merging processes, but only the first five papers were selected for `Hot' sequences to align with the length of `Top' and `Ordinary'. Next, for the 64,505 authors without hot streaks, we randomly chose five consecutive papers from their career data. If one or two of five papers were in the top 10\%, the sequence was classified as `Top', resulting in 25,227 cases. Otherwise, it was categorized as `Ordinary', accounting for 39,278 cases. Each sequence was tagged with the relative publication timing of its first paper, scaled between 0 and 1, to analyze within the same career stages. Various metrics were then calculated, based on the information from these five papers, such as author lists and topics.

\textbf{Team sizes.}
For the dataset used in this study, which includes 9,476,817 papers, the average number of authors per paper is 28, with a median of 4, a minimum of 1, and a maximum of 3,220. Notably, 88\% of the teams consist of fewer than ten members. Considering previous study\cite{wu2019large} and computation time, the characterisation in Figure \ref{fig:characterization} were performed for sequences with all five papers having fewer than ten authors each. 
We used a team size of ten as a benchmark for defining team sizes. `Big Project' is defined as a sequence of five papers where at least one paper includes more than ten authors (Fig. \ref{fig:characterization}e). The classification of whether a consecutive success involves large teams is based on whether at least one of the papers has more than ten authors (Fig. \ref{fig:explanation}a).

\textbf{Dense ties and loose ties.} To evaluate the collaboration patterns of a focal author's five consecutive papers, we examined their co-author list. If a co-author appears in all five papers, it signifies a strong collaboration, giving a maximum co-author appearance of five. If each paper has a different team without any overlaps, the value is one. We define sequences with a co-author in at least three out of five papers as having `dense ties'.

\textbf{Topic diversity.} The research topics of papers are calculated from co-citation network clustering of each author’s publication\cite{zeng2019increasing}, applying modularity maximization by Louvain method with resolution $=$ 1. 
For each consecutive paper, a topic diversity score is defined as the number of topics in these papers. The score is five when all papers are on different topics, while a score of one indicates all papers share the same topic. This score reflects the focal author's focus on a single or multiple research topics during that period.
The topic of each paper is calculated from clustering co-citation network on each researcher's publications \cite{zeng2019increasing} (Supplementary information \ref{subsec_sup:data_career}).

\textbf{New topics.} 
Within the targeted five papers sequence, we identify the most frequent topic $T_f$. If there is a tie for the highest frequency, the first topic appearing in the sequence is $T_f$. If $T_f$ does not exist in the author's earlier career, the attribute `New Topic' is true; if it was present before, it is False.


\section*{Acknowledgements}
The authors thank D. Wang, Y. Matsuo, and all members of our research group for their invaluable comments. 

\section*{Author contributions statement}
N.H. and T.M conceived the project;  N.H., T.M., Y.T. and A.K. collected data and performed analysis; N.H and T.M. wrote the manuscript, I.S supervised the project, all authors discussed results and edited the manuscript. Higashide and Miura have agreed that, due to their equal contribution as co-first authors, the order of their names can be interchanged when citing this paper or listing it in their CV.

\addtocontents{toc}{\protect\setcounter{tocdepth}{-1}}
\bibliography{references}
\addtocontents{toc}{\protect\setcounter{tocdepth}{2}}

\newpage
\section*{\huge Supplementary information}
\setcounter{figure}{0} 
\renewcommand{\thefigure}{S\arabic{figure}}
\renewcommand{\thetable}{S\arabic{table}}

\author[1,+*]{Noriyuki Higashide}
\author[1,+]{Takahiro Miura}
\author[1]{Yuta Tomokiyo}
\author[1]{Kimitala Asatani}
\author[1]{Ichiro Sakata}

\affil[1]{Department of Technology Management for Innovation, The University of Tokyo, Tokyo 113-8656, Japan}
\affil[+]{these authors contributed equally to this work}
\affil[*]{n.higashide@ipr-ctr.t.u-tokyo.ac.jp}


\tableofcontents

\section{Data description}
\label{sec_sup:data}

\subsection{Career sequence data}
\label{subsec_sup:data_career}
We prepared a home-built dataset of research papers based on Elsevier's Scopus Custom Data, encompassing all documents recorded between 1970 and 2021. 
As the most comprehensive database for abstracts and citations of peer-reviewed articles, Scopus contains over 73 million papers and 1.2 billion citations across a broad range of research fields and hence frequently utilized by researchers for bibliometric and citation analysis\cite{asatani2018cinet, baas2020scopus, miura2021sb, asatani2023}.
We focused on researchers who had sufficiently long careers to characterize timely-clustered success in life-long careers. Hence, we randomly extracted 100,000 authors who published more than 30 papers and 20 years of their whole careers from the dataset. To calculate citations within 10 years of publication, papers published from 2013 to 2021 are removed from the author information.
The historical order of publications was organized on a monthly basis, according to their publication dates.

To assess each paper's impact, we measured its citations within 10 years of publication, denoted as $C_{10}$ \cite{sinatora2016quantify}. 
We then adjusted for field-specific and annual variations by dividing it by the average citations for that field and year \cite{radicchi2008imp}. 
To obtain field information at the publication level, this study clustered the whole citation network in 2021 using the Leiden algorithm based on the Constants Potts Model ($\gamma=10^{-6}$)\cite{traag2011narrow, traag2019louvain}. Clusters with a size of less than $10^5$ papers were merged into other clusters with the strongest citation connections\cite{waltman2012new}, resulting in the identification of 66 distinct fields.
This re-scaling accounts for calibrating publication frequency and impacts across disciplines.

Each paper's topic is calculated by clustering the co-citation network within each career\cite{zeng2019increasing}. 
The topics were calculated by partitioning the co-citation network of papers published by each scientist using the Louvain method with a resolution parameter of $\gamma=1$.


\subsection{Potential limitation of datasets}
\label{subsec_sup:limitation_data}

The dataset used in this study has several limitations.
Firstly, this research targets researchers who have published 30 or more papers over a span of more than 20 years in Scopus, which may introduce survivorship bias and publication bias. 
For instance, researchers in fields such as law and philosophy, who may spend years producing a single book or publication, are less likely to be included in the dataset. 
Additionally, a plausible hypothesis that researchers who do not experience a hot streak may exit their careers earlier cannot be tested with this study. 
Moreover, publications in languages other than English are not recorded, which means that different outputs in other languages during periods of consecutive success might not be captured.
Scopus, by incorporating not only articles and conference proceedings but also book chapters, editorials, and letters, necessitates the consideration of diverse patterns of success that may vary according to document type.

Secondly, in the analysis of teams, the size may appear larger than in reality due to Elsevier's consolidation of group authors, which unfolds group contributions into individual ones. 
Group authors, such as the ``CONSORT Group," ``AHA Heart Failure Taskforce," or ``XY Workshop Members," attribute contributions to a group rather than individuals for outputs resulting from specific activities. 
Since the number of collaborators in such groups can be quite large, often in the dozens, this could potentially affect the perceived team size and density of ties.

Thirdly, although this analysis categorizes career stages as early-career and late-career within the dataset, the focus on publications from 1970 to 2012 means that individuals who began their careers before or continued their careers after this period might not have their actual career lengths accurately represented.

\newpage

\section{Probability of consecutive success}
\label{sec_sup:probability}

\subsection{Year distribution of U-shape success}
\label{subsec_sup:probability_year}

\begin{figure}[h!t]
    \centering
    \begin{minipage}[c]{.29\textwidth}
        \includegraphics[width=\textwidth]{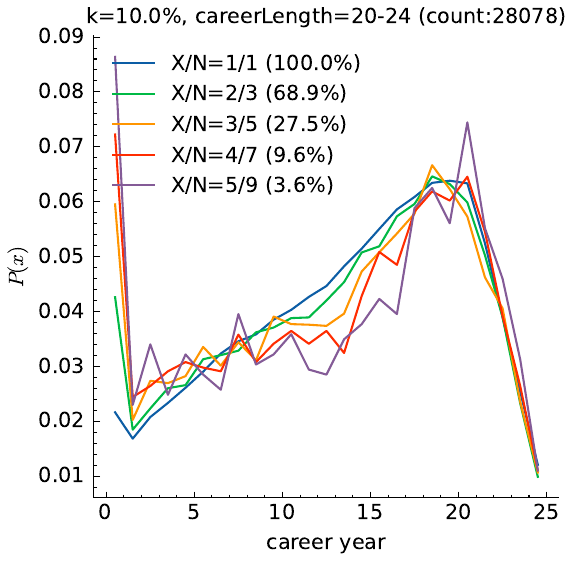}
    \end{minipage}
    \begin{minipage}[c]{.29\textwidth}
        \includegraphics[width=\textwidth]{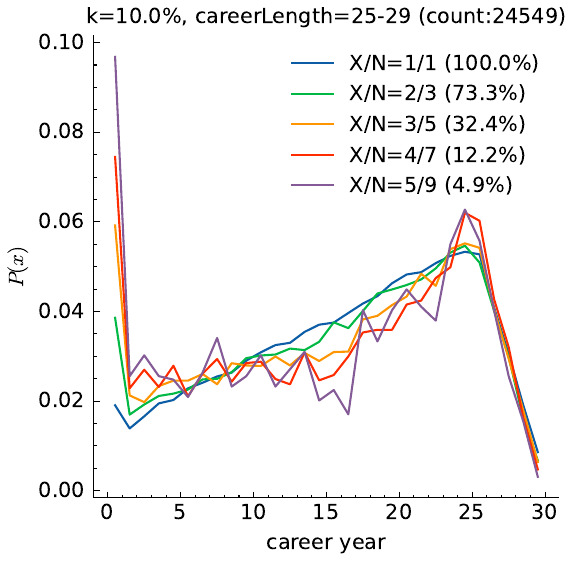}
    \end{minipage}
    \begin{minipage}[c]{.29\textwidth}
        \includegraphics[width=\textwidth]{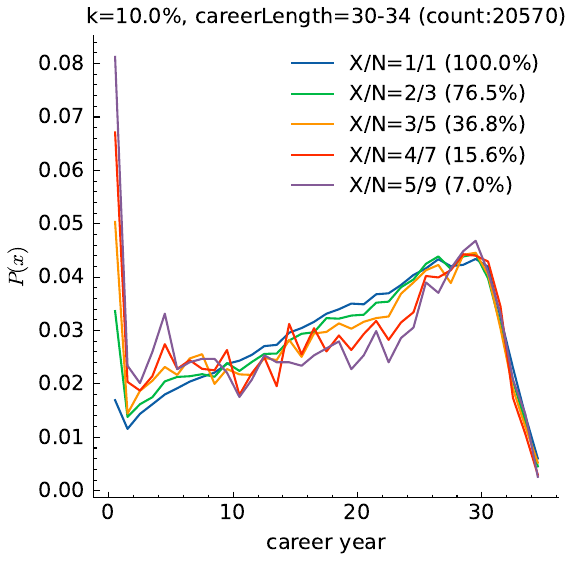}
    \end{minipage}
    
    \medskip 
    
    \begin{minipage}[c]{.29\textwidth}
        \includegraphics[width=\textwidth]{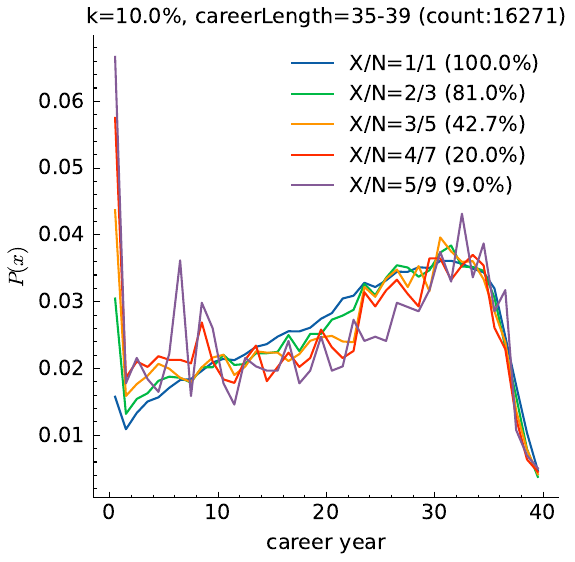}
    \end{minipage}
    \begin{minipage}[c]{.29\textwidth}
        \includegraphics[width=\textwidth]{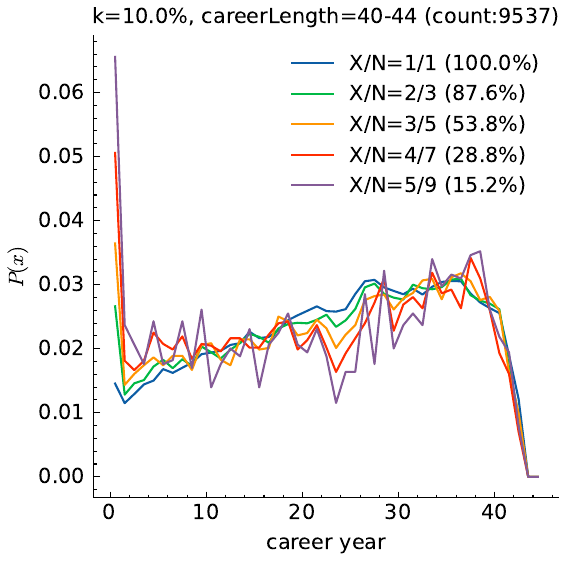}
    \end{minipage}
  \caption{The distribution of consecutive success among researchers with varying career lengths in five-year increments. Regardless of age, the probability of consecutive success declines from seven years after the career begins to 5-10 years before the career ends.}
  \label{fig_sup:u-shape_year}
\end{figure}

Although we have demonstrated the presence of a pitfall of consecutive successes in the mid-career phase on a relative career basis, it remains to be determined at what specific year of their career researchers are most susceptible to this phenomenon. 
Figure \ref{fig_sup:u-shape_year} illustrates the timing of consecutive success occurrences extracted using our model, translated into periods within a researcher's career. 
Researchers were divided into five groups based on career year length, and the distribution of single and consecutive successes was shown for each. It was found that consecutive successes become less likely approximately from seven years after the career begins to 5-10 years before the career ends. These timings likely correspond to the end of youth, the later stages of postdoctoral, and the periods heading towards retirement to complete careers. This indicates that researchers in their mid-careers may lack sufficient resources for adequate exploration and exploitation.
Note that the continuous increase in the number of single hits, which remain flat in terms of relative career from the early to the late stages, suggests that the number of publications tends to increase over time, possibly due to the utilization of collaborations, among other factors. 

\subsection{Time dependency of U-shape success}
\label{subsec_sup:probability_time}

The U-shape we observed might not just be about successes at the start or end of a scientist's career. 
Alterations in the dynamics of science can also serve as a potential confounder.
For example, if it is easier to publish high-impact papers around 2010 because of citation inflation\cite{petersen2019methods}, scientists at any time in their careers could have hot streaks recently.
To check this, we split scientists into five groups based on when they started their careers and how long their careers were. 
Then we checked if the U-shape still showed up.

\begin{figure}[h!t]
    \centering
    \begin{minipage}[c]{.19\textwidth}
        \includegraphics[width=\textwidth]{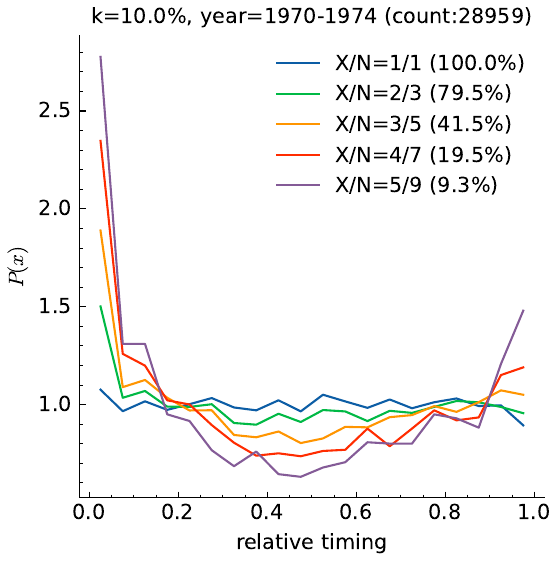}
    \end{minipage}
    \begin{minipage}[c]{.19\textwidth}
        \includegraphics[width=\textwidth]{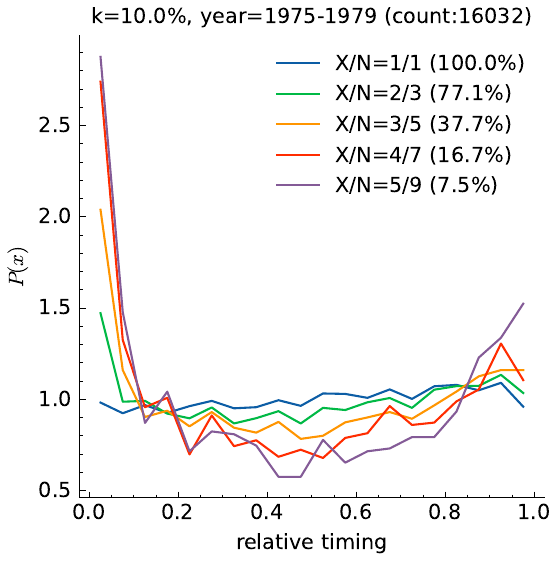}
    \end{minipage}
    \begin{minipage}[c]{.19\textwidth}
        \includegraphics[width=\textwidth]{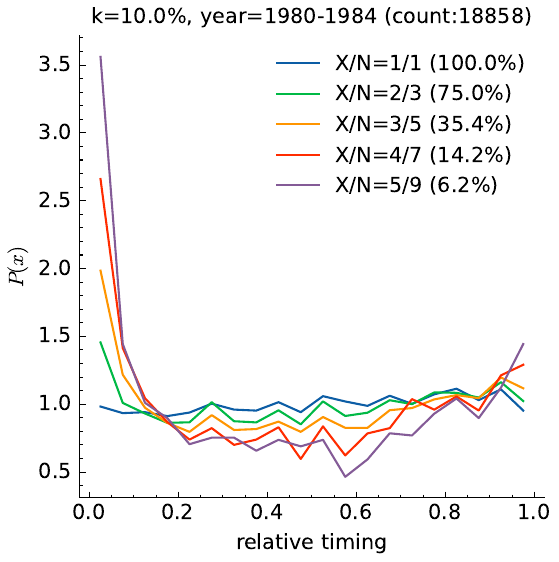}
    \end{minipage}
    \begin{minipage}[c]{.19\textwidth}
        \includegraphics[width=\textwidth]{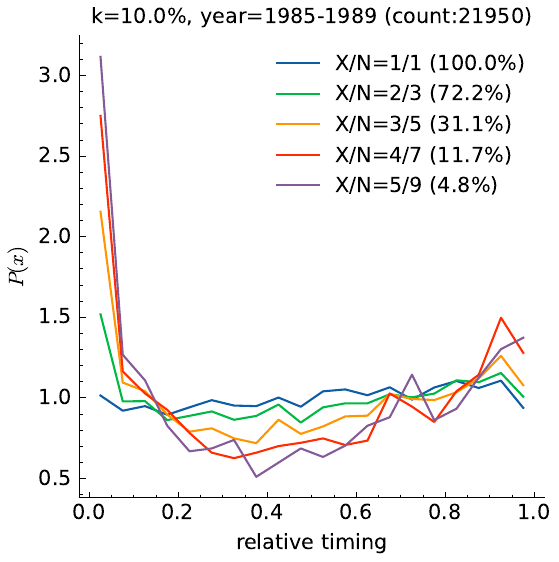}
    \end{minipage}
    \begin{minipage}[c]{.19\textwidth}
        \includegraphics[width=\textwidth]{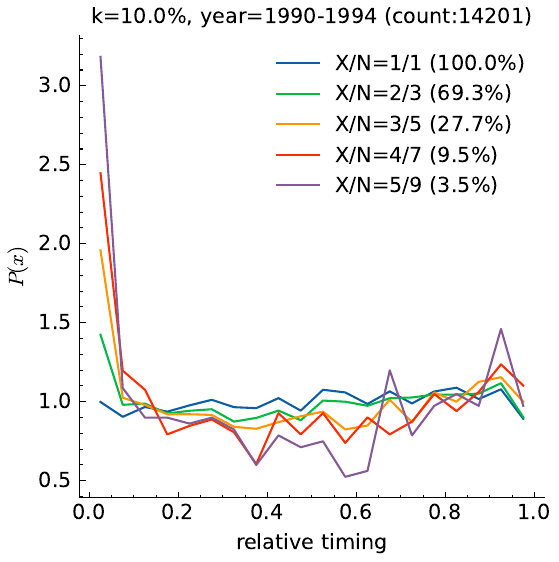}
    \end{minipage}
  \caption{The distribution of hot streaks among researchers with varying initiation periods of career in five years increments. Regardless of the ages, the pattern of consecutive success delineates a U-shaped curve.}
  \label{fig_sup:u-shape_begin}
\end{figure}

\begin{figure}[h!t]
    \centering
    \begin{minipage}[c]{.19\textwidth}
        \includegraphics[width=\textwidth]{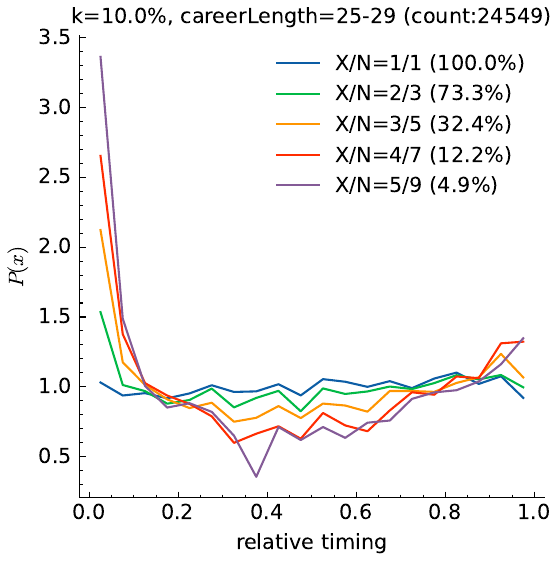}
    \end{minipage}
    \begin{minipage}[c]{.19\textwidth}
        \includegraphics[width=\textwidth]{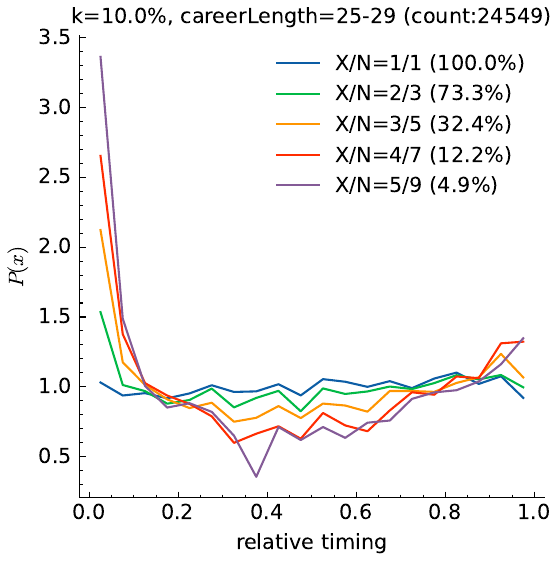}
    \end{minipage}
    \begin{minipage}[c]{.19\textwidth}
        \includegraphics[width=\textwidth]{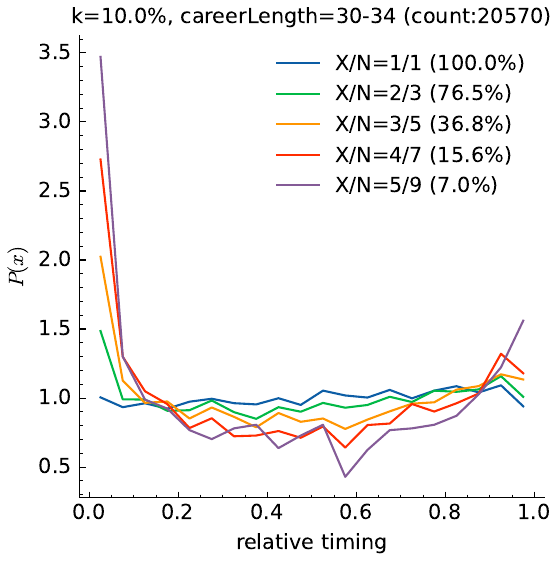}
    \end{minipage}
    \begin{minipage}[c]{.19\textwidth}
        \includegraphics[width=\textwidth]{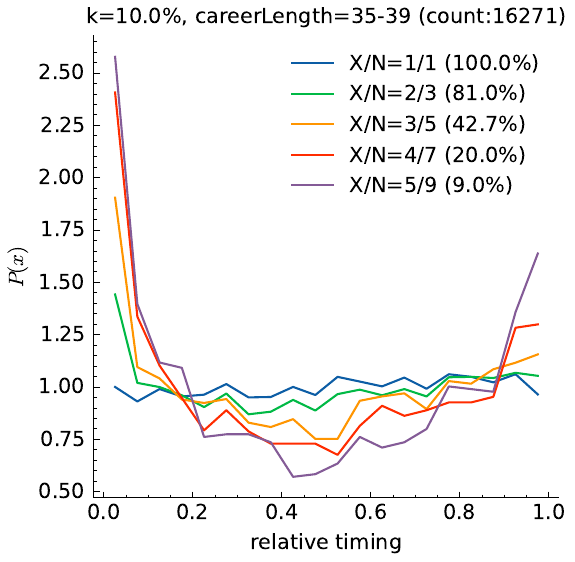}
    \end{minipage}
    \begin{minipage}[c]{.19\textwidth}
        \includegraphics[width=\textwidth]{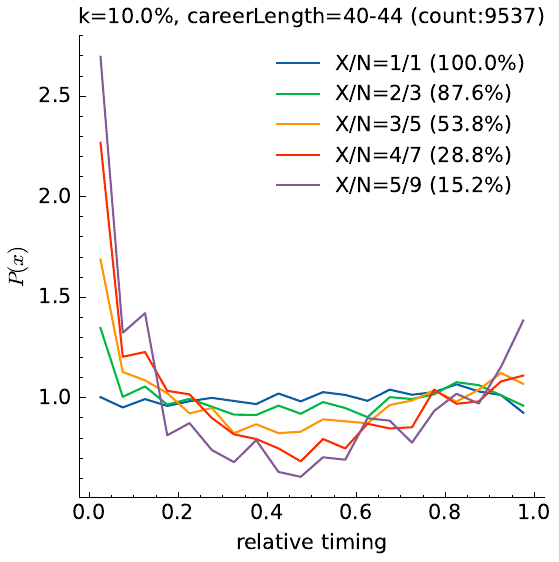}
    \end{minipage}
  \caption{The distribution of hot streaks among researchers with varying lengths of career in five-year increments. Researchers with careers of any duration exhibit a U-shaped curve in their pattern of consecutive success.}
  \label{fig_sup:u-shape_length}
\end{figure}

We found that no matter when a researcher started their career (Figure \ref{fig_sup:u-shape_begin}) or how long they've been working (Figure \ref{fig_sup:u-shape_length}), their single big successes happen randomly, but their hot streaks follow a U-shaped pattern. 
This is especially true for researchers with longer careers, as they publish more papers and thus have more chances for these hot streaks. For example, out of 9,537 researchers who have been publishing for over 40 years, 5,131 (about 53.8\%) had a hot streak where 3 out of 5 of their papers were in the top 10\%. 
This shows that the U-shaped pattern happens no matter the time period.

\subsection{Field dependency of U-shape success}
\label{subsec_sup:probability_field}

In scientific contribution, there are crudely two types of studies: empirical and theoretical.
Theoretical successes tend to appear early in a career, while empirical ones are more common later\cite{jones2014age}. 
To make sure the U-shaped pattern isn't just showing these different types of scientific contributions, we looked at whether the distribution of hot streaks varied across 53 fields when we identified the most common topic during each hot streak as the topic of that streak.

\begin{figure}[h!tbp]
    \centering
    \includegraphics[width=.7\linewidth]{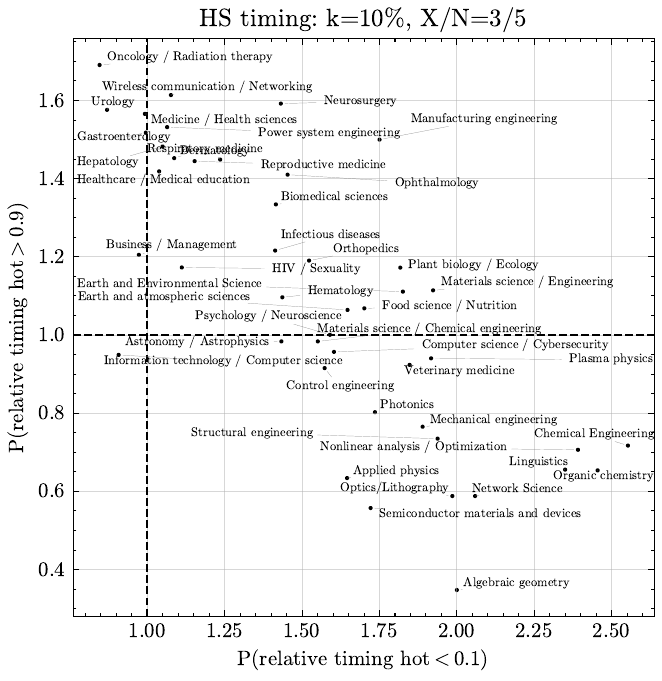}
    \caption{The distribution of U-shaped patterns in success across various fields (defined by k=10\%, X/N=3/5). Each hot streak(HS) is categorized by its most common topic, and the graph plots the proportion of hot streaks occurring either early (probability P(x)<0.1) or late (probability P(x)>0.9) in a career. The thick dotted line indicates P(x)=1, with fields located in the upper right quadrant demonstrating a U-shaped pattern, signifying a higher occurrence of both early and late career peaks in success. The names of the fields were assigned based on the highest TF-IDF values of keywords in papers.}
    \label{fig_sup:ushape_field}
\end{figure}

We made a scatter plot in Figure \ref{fig_sup:ushape_field}. 
It shows if the probability distribution of hot streaks in each field is more than 1 either early in the career (relative timing below 0.1) or late (above 0.9).
Out of 45 fields where more than 100 people experienced a hot streak (defined as k=10, X=3, N=5), 21 fields showed a U-shaped pattern with both early and late hot streaks. 
However, 18 fields had more early but fewer late hot streaks, like in Algebraic Geometry or Semiconductor materials and devices, which include both theoretical and empirical studies. 
Overall, early career hot streaks are robust, but whether one experiences late career hot streaks seems to depend on the field.
\newpage

\newpage
\section{Relation with the existing model}
\label{sec_sup:liu_model}

\subsection{Empirical measurements}
Hot streaks are a phenomenon in which high-impact papers are concentrated at a time throughout a scientist's career \cite{liu2018hot}. Although there are minor data differences, we were able to confirm that the hot streak phenomenon is reproduced in our Scopus dataset.

\begin{figure}[h!tb]
    \centering
    \includegraphics[width=\linewidth]{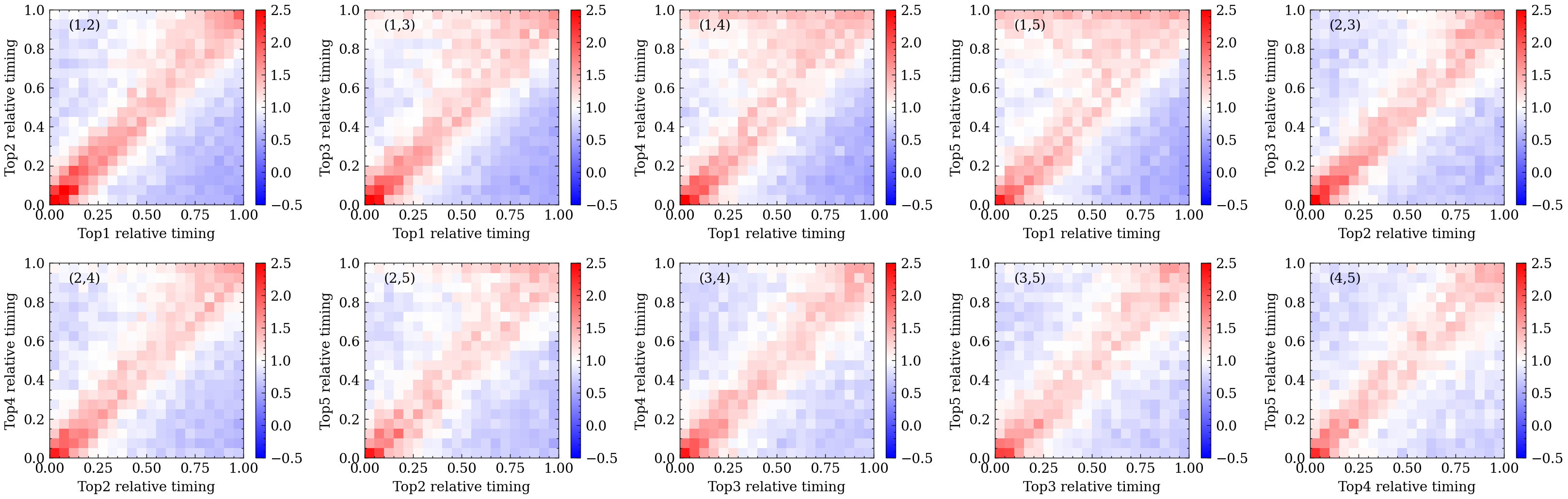}
    \caption{Distribution of normalized joint probabilities $P(x_1,x_2)/P(x_1)P(x_2)$ for the top-1 and 2 high-impact work, $x1$ and $x2$. Plotted 10 combinations of top 1-5 with the bins size of 20. The diagonal pattern indicates that hits tend to occur relatively consecutively. Relative timing is the $N$th paper the researcher has published divided by the total number $N_T$. For example, the timing of the 8th paper published by an author who has written 80 papers would be 0.1.}
    \label{fig:joint_prob_maps}
\end{figure}

\begin{figure}[h!tb]
    \centering
    \includegraphics[width=\linewidth]{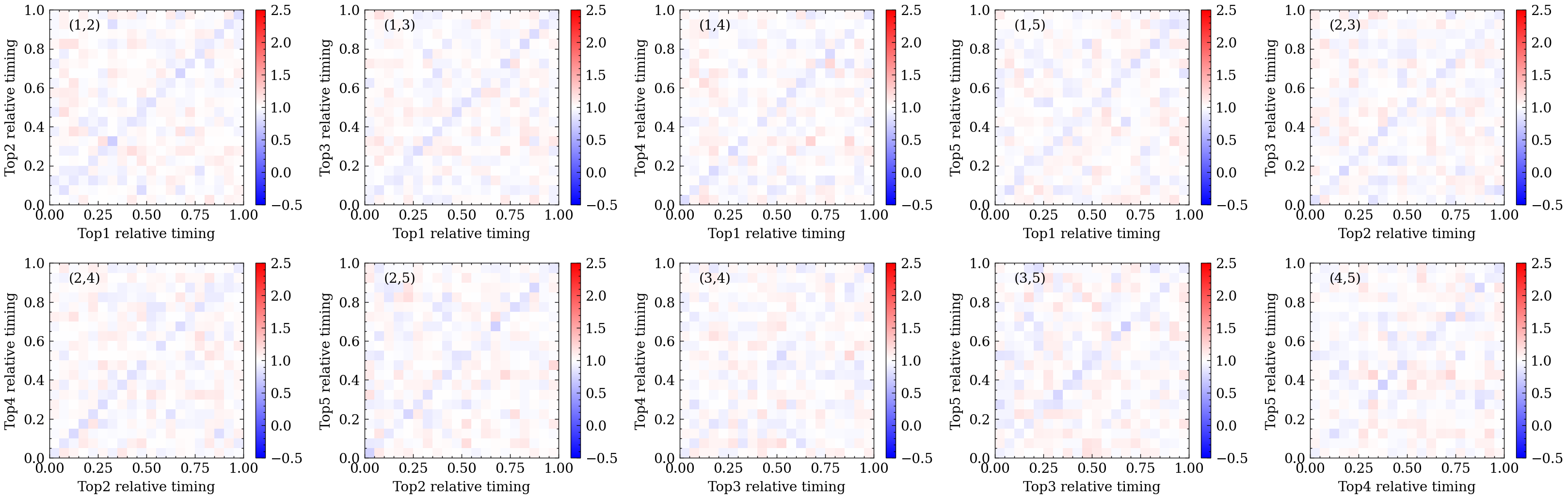}
    \caption{Distribution of normalized joint probabilities of the $x1$ and $x2$ for the career sequence data shuffled in order. The diagonal pattern has disappeared.}
    \label{fig:joint_prob_maps_shuffle}
\end{figure}

We began by investigating the timing of the five most impactful works produced in each career. 
As confirmed in the original paper, we confirmed that the probabilities of when the top-impact works occur to each other are correlated. The normalized joint probability of the highest and second highest-impact work is calculated and displayed as a heatmap. Similar calculations were performed for each of the top 1-5 pairs and 10 combinations, and for all pairs, a pattern of high joint probability was observed on the diagonal (Fig. \ref{fig:joint_prob_maps}). This indicates that the timing of the appearance of the top 1-5 in the careers is correlated and that they tend to occur consecutively. This diagonal line disappears when the order of the carrier series data is shuffled, indicating that this tendency is unique to the actual data (Fig. \ref{fig:joint_prob_maps_shuffle}).

\begin{figure}[h!tb]
    \centering
    \includegraphics[width=\linewidth]{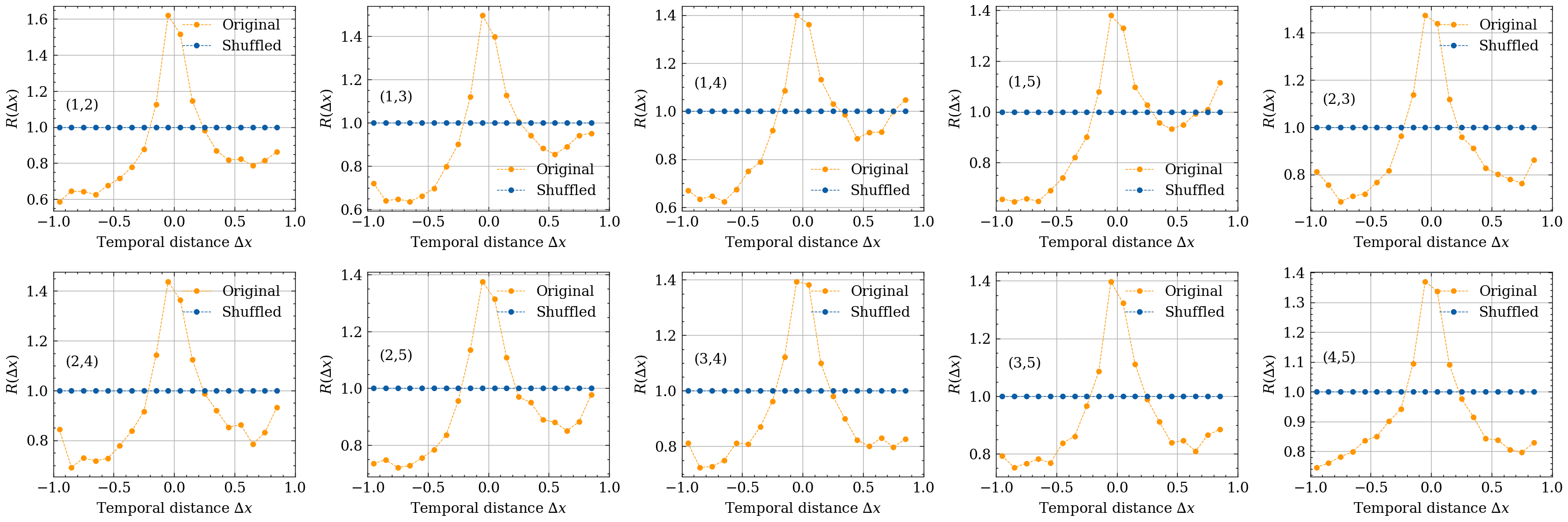}
    \caption{Normalized distribution of the difference in the relative timing of the top 1,2,3,4,5 papers. All are divided by the distribution of shuffled data. The yellow dots show the actual data and the blue ones show the shuffled data.}
    \label{fig:temporal_distance}
\end{figure}

\begin{figure}[h!tb]
    \centering
    \includegraphics[width=0.8\linewidth]{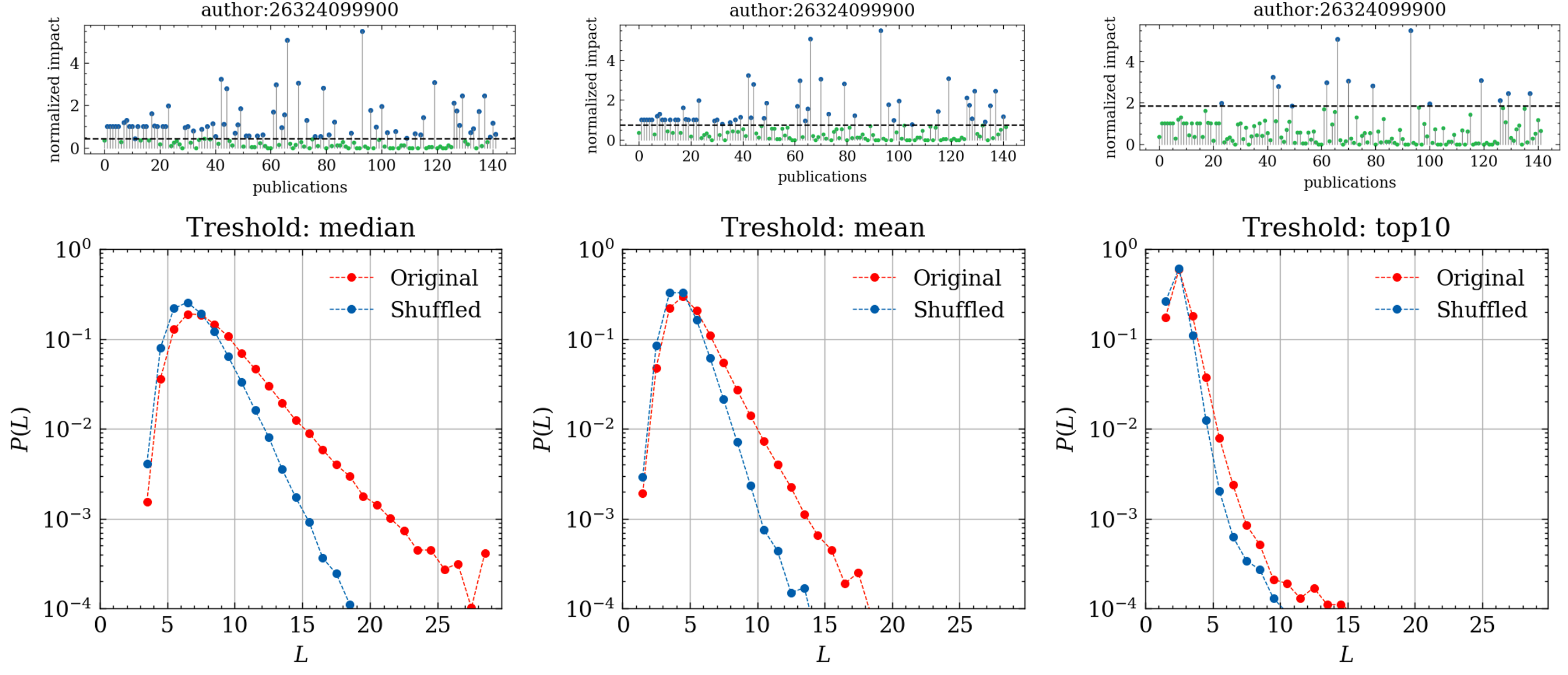}
    \caption{(Top row) Career sequence data for a scientist. The dotted line is the threshold, and higher-impact jobs are indicated by blue dots. From left to right: median, average, and top 10\% thresholds. (Bottom row) Of works higher than the thresholds, the largest consecutive length is $L$, and its distribution is shown in red. Blue is for shuffled data. The higher the threshold, the shorter the $L$ goes, but all of them are longer than the shuffle.}
    \label{fig:streak_length}
\end{figure}

We defined temporal distance ${\Delta}x$ as the difference in the relative timing of the top 1-5 papers to appear and divided the distribution $P({\Delta}x)$ by the distribution of the shuffled data $P_s({\Delta}x)$, we observed distribution $R({\Delta}x)=P({\Delta}x)/P_s({\Delta}x)$ (Fig. \ref{fig:temporal_distance}). If $R({\Delta}x)$ is higher than 1, it is more likely to occur than in the random case. The top 1-5 hits all have high values around 0 confirming that they are also more likely to occur at near time.

We replicated the original paper not only for timing but also for the length that high-impact works continue. We set a certain threshold and observed the distribution of the maximum length $L$, of the consecutive periods of higher-impact work, which tends to be longer than that of the shuffled data (Fig. \ref{fig:streak_length}). This indicates that the actual data tends to be more likely to be followed by a series of higher-impact papers than the shuffled data. To this end, hot streaks where high-impact papers are consecutive in time are also observed in our Scopus data set.

\subsection{Revisiting hot-streak model}

\begin{figure}[h!tb]
    \centering
    \includegraphics[width=0.85\linewidth]{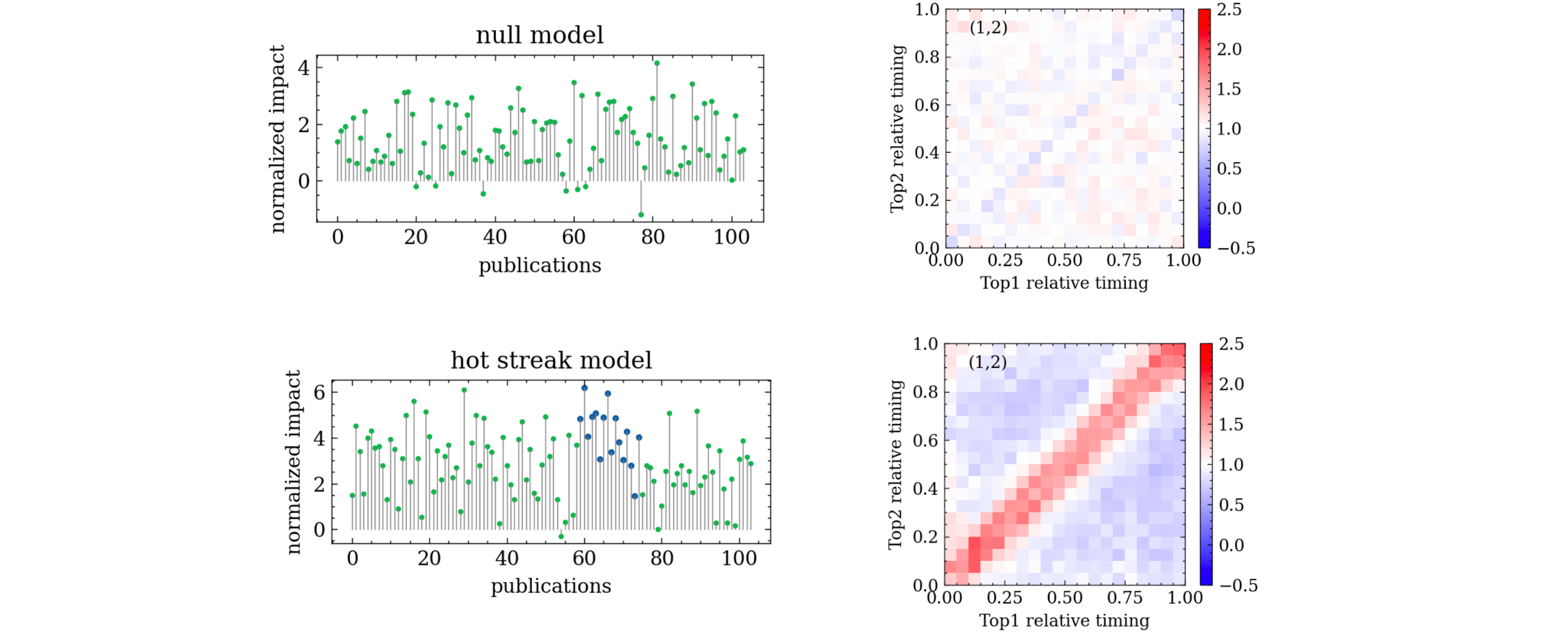}
    \caption{We generated two types of sequence datasets following the null model and hot-streak model respectively. As shown in previous research, we set ${\Gamma}_0$ to represent the average impact level of each researcher, with ${\sigma}) = 1.0$, and ${\tau}_H$ as 4 years.}
    \label{fig:generative_model}
\end{figure}

\begin{figure}[h!tb]
    \centering
    \includegraphics[width=1.0\linewidth]{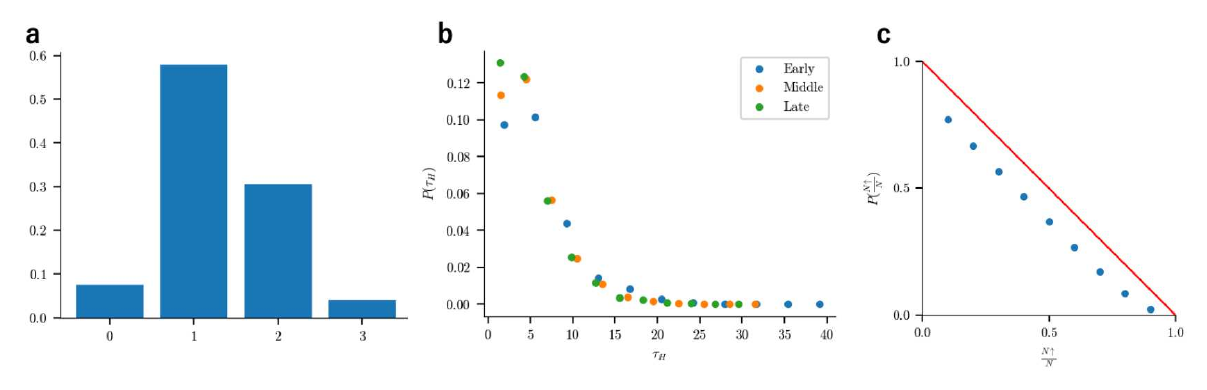}
    \caption{Characteristics of reproduced hot-streak model. \textbf{a}, Histogram of the number of hot streaks in a career. \textbf{b}, The distributions of durations of hot streaks $P(\tau_H)$. \textbf{c}, The cumulative distributions of the onset timing $P(N_{\uparrow}/N)$.}
    \label{fig:repro_fig2}
\end{figure}

To explain the observed temporal regularities in career sequence data, the previous research has proposed a simple model \cite{liu2018hot}. First, they introduced a null model, where each work is randomly selected from a normal distribution with parameters $\mathcal{N}({\Gamma}_0, {\sigma})$, where ${\Gamma}_0$ represents the individual researcher's specific parameter of impact level, and ${\sigma}$ represents the variability in impact. With these two parameters and the total number of papers $N_T$, we can generate career sequence data following the null model. Next, they proposed hot-streak model, where ${\Gamma}_0$ increases to ${\Gamma}_H = {\Gamma}_0 + 1.0$ during a certain period ${\tau}_H$ randomly chosen within the career. By considering this simple assumption, they could reproduce patterns in actual data.

Using our dataset, we generated 100,000 sequence data following both the null model and the hot-streak model (Fig. \ref{fig:generative_model}). Interestingly, in the null model, the observed patterns in the simultaneous probability map did not exhibit diagonal patterns, while in the hot-streak model, diagonal patterns were observed. This indicates that the simple assumption reproduced the temporal proximity of success in the actual data, as reported in the prior study.

To apply the hot-streak model to actual data, we fitted a specific piecewise function to the moving average ${\Gamma}(N)$ of the career sequence data. This function resembled a square wave pulse and could have up to three rising edges, characterized by ten parameters:$base$, $hot_{1}$, $N_{1\uparrow}$, $N_{1\downarrow}$, $hot_{2}$, $N_{2\uparrow}$, $N_{2\downarrow}$, $hot_{3}$, $N_{3\uparrow}$, $N_{3\downarrow}$. Here, consistent with the previous research, ${\Gamma}(N)$ denotes the moving average over a window size ${\Delta}N$, determined as 10\% of the total number of works created, $N_T$ on an individual basis. It is important to note that ${\Gamma}(N)$ does not reflect the impact of the N-th work on the individual. Instead, it summarizes the average performance of works created before and after the N-th piece. To ensure a robust statistical representation, we adopted the definition ${\Delta}N = max(5, 0.1N_T)$ from the previous research, ensuring sufficient statistics when calculating ${\Gamma}(N)$. Additionally, to prevent overfitting, we introduced an L1 regularization term on the difference between $base$ and each $hot_1$, $hot_2$, $hot_3$. 20 random initial values are prepared for the other parameters, adopting the one with the smallest squared error.

To examine the characteristics of fitting with the hot-streak model, we quantified the number of occurrences of hot streaks in a career. Consistent with previous research, Fig. \ref{fig:repro_fig2}a shows that one occurrence was most common, followed by two and zero occurrences, with three occurrences being the least common. Fig. \ref{fig:repro_fig2}b indicates that the distribution of hot streak lengths in years, remained constant across different career stages, averaging 4.1 years almost similar to previous studies. Lastly, we aggregated the start timing of hot streak in a career (Fig. \ref{fig:repro_fig2}c). Hot streaks were slightly more common in the early stages of the carriers, while the probability was about the same in the other periods. In previous studies, the probability of hot streak occurrence was equal for all carriers, which is a slight difference from the model reproduced here. These results indicate that we have reproduced the hot-streak model of a certain quality.


\subsection{Dependency of the hot-streak model on window size of \texorpdfstring{$\Gamma(N)$}{Gamma(N)}}
\label{subsec_sup:hs-model-limits}

\begin{figure}[h!tb]
    \centering
    \includegraphics[width=\linewidth]{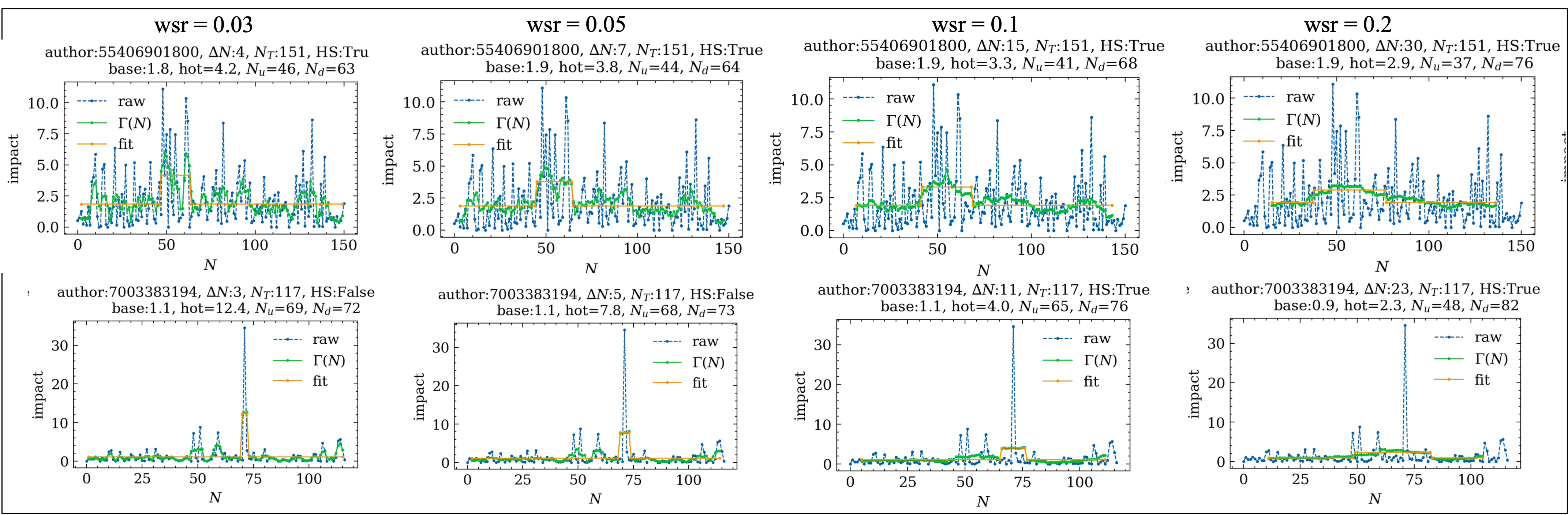}
    \caption{For two researchers, the moving average ${\Gamma}(N)$ (green) was calculated from the sequence data (blue), and a function assuming a single hot streak was fitted (orange). It can be observed that for the same data, as the window size increases,  ${\Gamma}(N)$ becomes smoother and the fit goes lower and longer. The fitting was done by exploring all possible regions of four parameters under certain constraints, with the least squared error.}
    \label{fig:figures/hotstreak_fitting.png}
\end{figure}

In the context of successfully replicating the hot-streak model, we focused our attention on the window size that serves to smooth out the impact and found that the shape of ${\Gamma}(N)$ changes depending on the window size of the moving average ${\Delta}N$. With the definition of ${\Delta}N = max(5, wsr*N_T)$, in which $wsr$ represents window size ratio among each career, we then calculated ${\Gamma}(N)$ for career sequence data at four different $wsr$ values. Looking at the two sequences as examples, the shape of ${\Gamma}(N)$ becomes smoother as $wsr$ increases (Fig. \ref{fig:figures/hotstreak_fitting.png}). 

\begin{figure}[h!tb]
    \centering
    \includegraphics[width=0.7\linewidth]{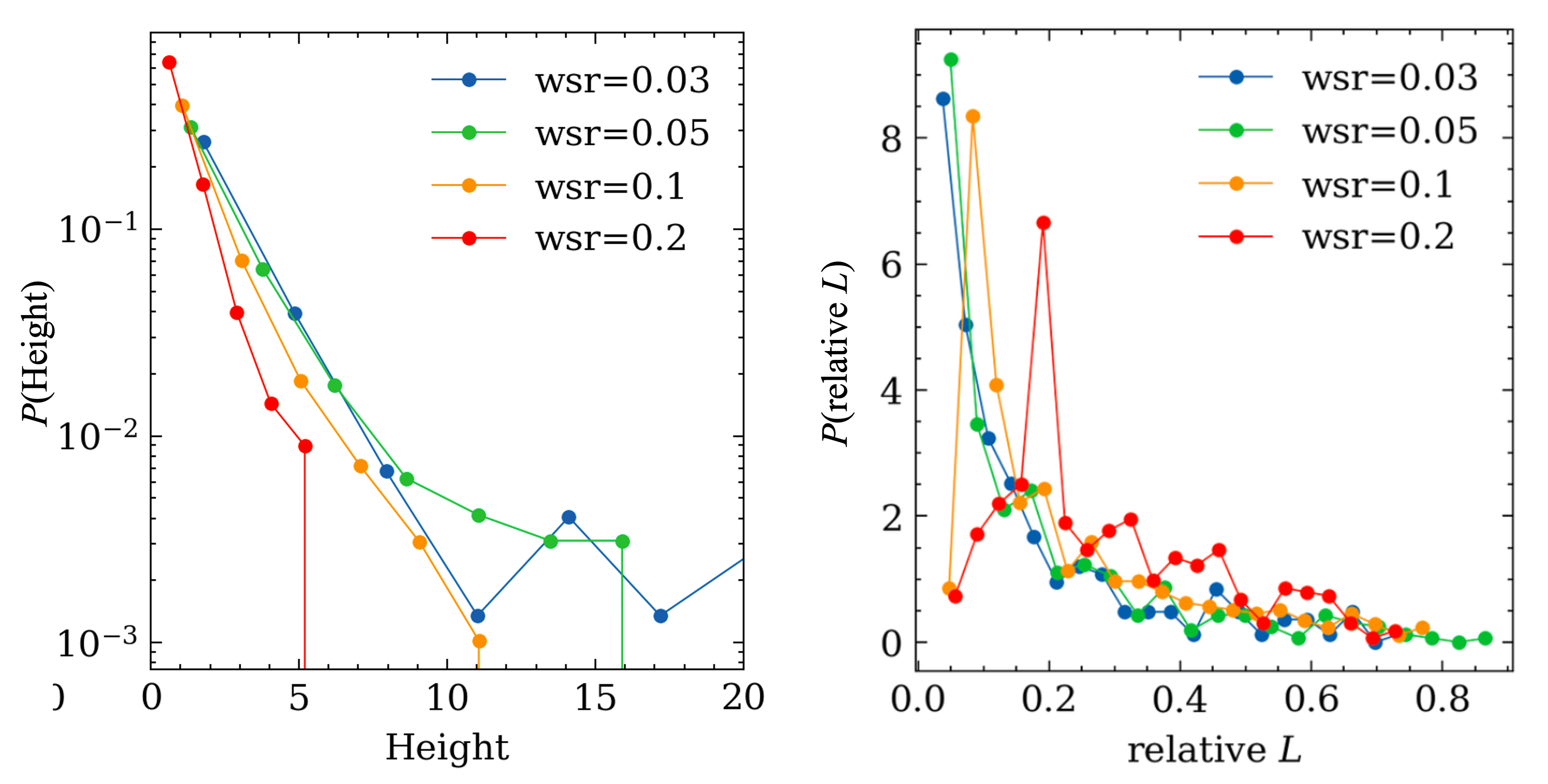}
    \caption{The distribution of parameters for the simple version of hot-streak model fitted to 500 researchers. As the window size increases, the height of the hot streak decreases, and its length increases.
    The height is defined as $hot - base$ and the length $realtive L$ is the normalized length $(N_{d} - N_{u})/N_T$.}
    \label{fig:hotstreak_fitting_param_depend}
\end{figure}

To test the effect of $wsr$, we fitted a simple version of hot-streak model over the several ${\Gamma}(N)$ with four different $wsr$. The simple version assumes only one raise thus it has four parameters of $hot$, $base$, $N_{u}$ and $N_{d}$. In Fig. \ref{fig:hotstreak_fitting_param_depend}, the impact elevation $Height$, calculated from $hot - base$, declines as $wsr$ increases and the length of the hot streak, indicated by the difference between $N_{d}$ and $N_{u}$, increases. This trend is confirmed by fitting the parameters for 500 researchers. 

Notably, the distribution of the length of hot streaks $realtive\ L$ peaks at the same length of $wsr$. 
This can be understood more clearly by examining the lower row of Figure \ref{fig:figures/hotstreak_fitting.png}. When career sequence data with a single outstanding impact are converted to ${\Gamma}(N)$, a shape resembling a plateau around the maximum impact emerges, whose length almost matches $wsr*N_T$, as intuitively obvious from the definition of the moving average. Fitting such ${\Gamma}(N)$ results in tracing a square-wave-pulse shape, suggesting a potential correlation between the relative length of the hot streak $L$ and $wsr$. In such cases, the model detects a single significant success rather than consecutive successes. Given its current definition, the hot-streak model cannot prevent such false-positive detections. Therefore, there may be room for improvement in accurately detecting consecutive successes using the hot-streak model.

Indeed, there are cases where consecutive successes are detected, indicated by periods where the overall impact increases as shown in the upper row of Figure \ref{fig:figures/hotstreak_fitting.png}. The proposed hot-streak model undeniably contributes by capturing a common feature even among film directors and painters, which is the occurrence of consecutive hits. However, the main claims such as 1) 90\% of researchers experience a hot streak, 2) hot streaks can occur at any time throughout a career, and 3) the average length of a hot streak is about four years, does not fully capture general trends applicable in all cases. 
This is because these are based on the distribution of fitting parameters when the window size for the moving average is arbitrarily assumed to be 10\% of the career length. 

\subsection{The hot-streak model and our data-centric detection}

\begin{figure}[h!tb]
    \centering
    \includegraphics[width=\linewidth]{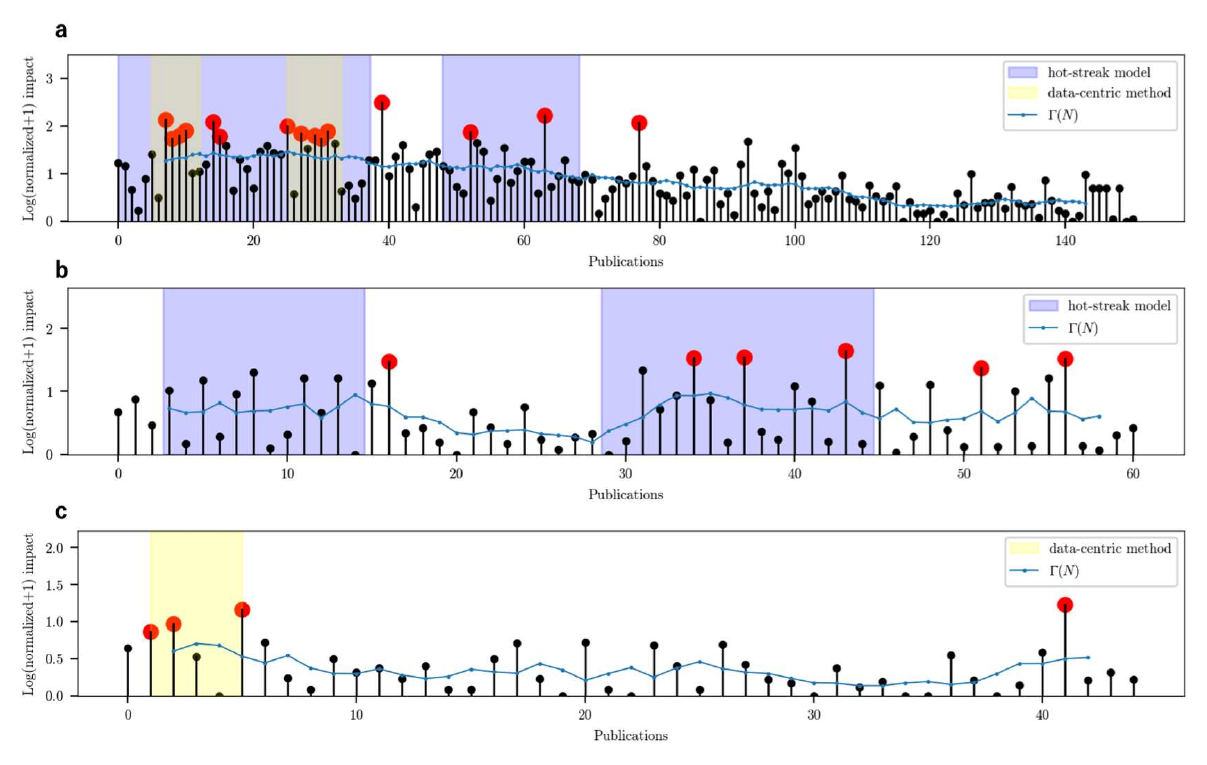}
    \caption{Differences in hot streak extraction trends between the hot-streak model and our data-centric approach. Red circles indicate the top-10\% works in a career. \textbf{a}, When the timing of the hot streak was common. \textbf{b}, Cases in which the onset timing is not common and more than one hot streak was extracted by the hot-streak model. \textbf{c}, Cases where the onset timing was not common and one or more of them were extracted by the data-centric detection.}
    \label{fig:sample_confusion_matrix}
\end{figure}

To evaluate the correspondence between the replicated hot-streak model and our method, data-centric consecutive success detection, we examined the extent of alignment in the onset of hot streak periods. We normalized career timelines from 0 to 1 and defined the onset of a hot streak within a tolerance range of ±0.1 as a match. The comparison results are presented in a confusion matrix (Table. \ref{table_sup:confusion_matrix}). 
Out of the consecutive success periods identified using our data-centric method, only 47\% are also detected by the hot-streak model. This means our method overlooks roughly half of the periods caught by the model. 
Conversely, of the consecutive successes detected by the hot-streak model, only 19\% are identified by the data-centric method. 
Given this, the hot-streak model tends to identify a significant portion of the periods as ``success streaks" even when consecutive success of top-10\% works is not occurring. Even in periods of consecutive success, it fails to recognize them about half the time (Fig. \ref{fig:sample_confusion_matrix}). In this study, we demonstrate the validity of our approach.

\begin{figure}[h!tb]
    \centering
    \includegraphics[width=1.0\linewidth]{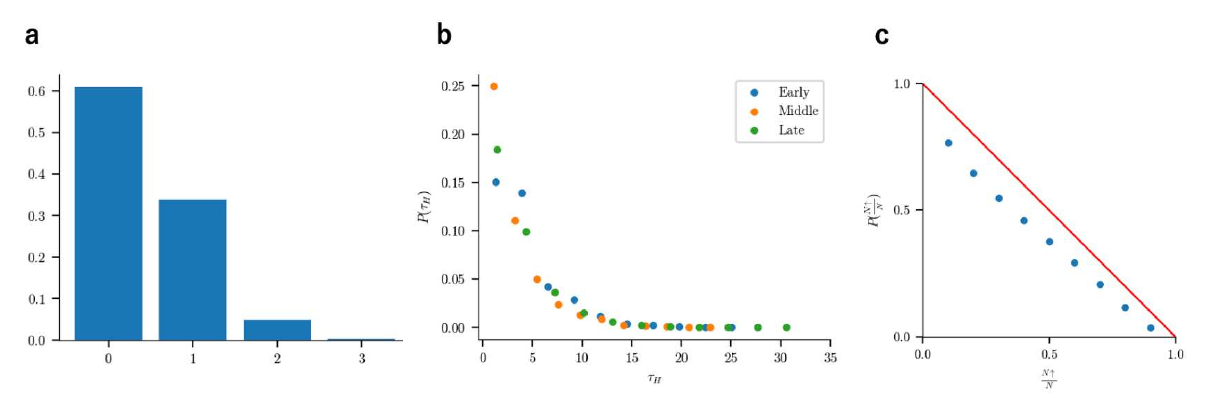}
    \caption{Characteristics of reproduced hot-streak model with raw-impact arrays. \textbf{a}, Histogram of the number of hot streaks in a career. \textbf{b},The distributions of durations of hot streaks $P(\tau H)$. \textbf{c}, The cumulative distributions the start of a hot streak.}
    \label{fig_sup:liu_fit_raw}
\end{figure}

When employing the hot-streak model on sequences of raw impacts without applying a moving average, approximately 30\% of researchers are found to experience at least one hot streak, while less than 10\% undergo multiple hot streaks (Fig. \ref{fig_sup:liu_fit_raw}a). 
This closely aligns with our model's ratio of X/N=3/5 and suggests that the 90\% hot streaks identified in prior studies may be artifacts of high impacts being smoothed out by moving averages. 
Furthermore, while the average duration of hot streaks in models with a set window size is around four years, the variation is significantly greater in hot-streak models fitted to raw data (Fig. \ref{fig_sup:liu_fit_raw}b). 
Comparing our method to the raw data fitting hot-streak model reveals that the latter does not fit well with non-consecutive successes. 
Comparing Tables \ref{table_sup:confusion_matrix} and \ref{table_sup:confusion_matrix_raw}, the number of hot streaks detected only by our data-centric method is fewer in the raw data fit. The count of hot streaks identified solely by the hot-streak model has increased.
However, similar to the moving average, the distribution of the onset of hot streaks appears nearly flat(Figure \ref{fig_sup:liu_fit_raw}c).

As limitations, due to computational constraints, the fitting was only performed on 500 individuals, and the method only considers up to a single hot streak, thus it is not a complete replication experiment. In the future, following previous studies, a method to fit up to three hot streaks will be applied to sufficient data of 100,000 individuals to observe more precise parameter distribution.

Our data-centric detection approach is simpler and directly engages with raw data, enabling a more accurate capture of real-world phenomena. The hot-streak model, on the other hand, works with career sequence data smoothed by moving averages. This model doesn't always successfully identify sequences of good work in a career, specifically the concentration of top-10\% achievements, with only a 50\% probability.

\begin{table}[hbtp]
\begin{minipage}[t]{.5\textwidth}
 \centering
 \caption{The number of hot streaks detected \\ by hot-streak model (n=100,000)}
 \label{table_sup:confusion_matrix}
 \renewcommand{\arraystretch}{1.5} 
\begin{tabular}{cc|cc}
\multicolumn{2}{c}{}
            &   \multicolumn{2}{c}{\small Data-centric detection} \\
    \multirow{2}{*}{\rotatebox[origin=c]{90}{\small Hot-streak model}}
    &       &   Positive &   Negative              \\
    \cline{2-4}
    & Positive   & 24,030   & 104,310                 \\
    & Negative    &  22,293   & -                \\
    \cline{2-4}
\end{tabular}
\end{minipage}
\hfill
\begin{minipage}[t]{.5\textwidth}
 \centering
 \caption{The number of hot streaks detected \\ by hot-streak model for raw data (n=100,000)}
 \label{table_sup:confusion_matrix_raw}
 \renewcommand{\arraystretch}{1.5} 
\begin{tabular}{cc|cc}
\multicolumn{2}{c}{}
            &   \multicolumn{2}{c}{\small Data-centric detection} \\
    \multirow{2}{*}{\rotatebox[origin=c]{90}{\begin{tabular}{@{}c@{}}Hot-streak model \\for raw data\end{tabular}}}
    &       &   Positive &   Negative              \\
    \cline{2-4}
    & Positive   & 12,667   & 30,531                 \\
    & Negative    &  33,916  & -                \\
    \cline{2-4}
\end{tabular}
\end{minipage}
\hfill
\end{table}

\clearpage

\section{Collaboration patterns}
\label{sec_sup:team}

\subsection{Team sizes}
\begin{figure}[h!tp]
    \centering
    \includegraphics[width=0.3\linewidth]{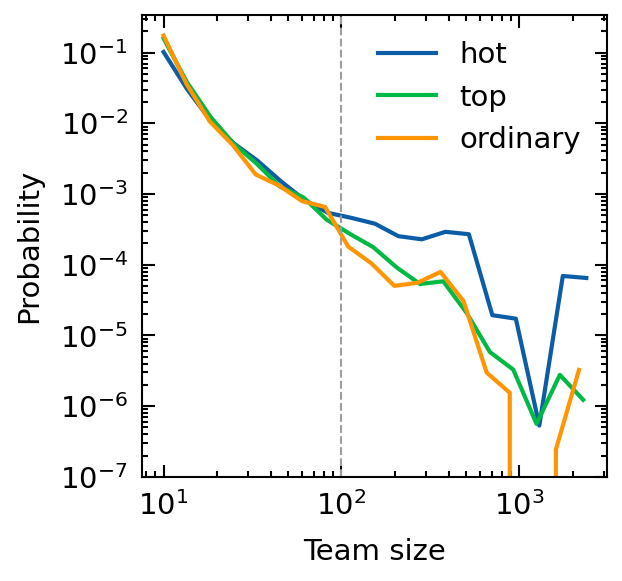}
    \caption{Histograms of team sizes for each sequence. The dashed line is on 100 of team size. `Hot' sequences have greater proportion of big project.}
    \label{fig_sup:teamsize_prob}
\end{figure}

In sequences experiencing consecutive successes, `Hot' sequences, team sizes tend to be larger, with a higher proportion of large projects involving more than 10 members (Fig. \ref{fig:characterization}d,e). Histograms of team sizes for each sequence show similar proportions for teams up to 100 members, but `Hot' sequences have a higher proportion of very large projects with over 100 members (Fig. \ref{fig_sup:teamsize_prob}). This indicates that large-scale projects often publish a cluster of papers at once, likely contributing to the top 10\% high-impact papers in a career.

\subsection{Examining if early and late hot streaks come from the same work}
We hypothesized that consecutive successes in early and late career stages may originate from the same teams. For instance, a researcher who had multiple hits as a student or postdoc in their early career might be linked to a mentor or principal investigator who appears as the last author during the same period. Therefore, these early-career mentors, often senior researchers later in their careers, could mirror the later career successes. To test this hypothesis, 20,000 papers were randomly sampled from early and late career consecutive success sequences, defined as having a relative career timing less than 0.2 for early and greater than 0.8 for late stages. If these overlap, it indicates the authors experienced the consecutive successes at the same time in their each career stages, early and late. The overlap was measured using the Jaccard index, which was 0.002. For randomly extracted paper sets, regardless of career stage, the Jaccard coefficient was 0.1. This suggests that `Hot' papers in early and late stages are more likely to written by different teams, strongly supporting the notion that early and late career successes are qualitatively distinct. Notably, only 1.7\% of individuals, 474 in total, experience consecutive successes in both early and late stages of their career lifecycle.

\clearpage

\subsection{Disruption index of hot works and works of who experience hot streak}

\begin{figure}[h!tp]
    \centering
    \includegraphics[width=0.3\linewidth]{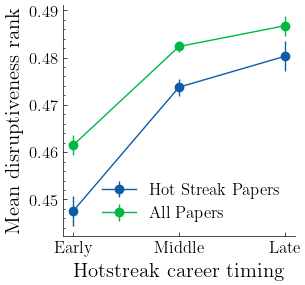}
    \caption{The blue line depicts the disruptiveness rank\cite{wu2019large} (D) of hot streak papers during the early, middle, and late periods. The green line represents the D rank of all papers authored by scientists experiencing early, middle, and late hot streaks. The error bar for both series indicates a 95\% confidence interval. The rank of D is determined within groupings that are classified according to year (in intervals of 5 years), cited count (divided into 5 percentiles), and the field (calculated in \ref{subsec_sup:data_career}).}
    \label{fig_sup:D_period}
\end{figure}

Our findings indicate that scientists belonging to the early hot-streak group  (experience hot streak up to 20\% of the career timing) tend to experience less disruption in their scholarly careers compared to those in the middle (20-60\% of the career timing), late (80\% onward of the career timing) group. Furthermore, even in their hot streak papers, the early group displays low disruptiveness when compared to the other groups. These results suggest that early hot streaks are characterized by a limited level of topic diversity (see Figure \ref{fig_sup:D_period}). Scientists in this group tend to concentrate on developing their research and maintain a consistent style throughout their careers. Conversely, the late hot-streak group demonstrates relatively high levels of disruptiveness, despite also exhibiting low topic diversity (see Figure \ref{fig:characterization}h). This can be attributed to their involvement in or leadership of large research teams and the presence of weak ties (see Figures \ref{fig:characterization}d, f). These findings imply that they engage in research that incorporates diverse ideas and collaborations, thereby avoiding research with low levels of disruptiveness.

\end{document}